\documentclass[%
 superscriptaddress,
 	preprint,
 showpacs,
	preprintnumbers,
 nofootinbib,
 amsmath,
 aps,
	prd,
]{revtex4-1}

	\usepackage{fullpage}
	\usepackage[toc,page]{appendix}
	\usepackage{amsmath}
	\usepackage{amssymb}
	\usepackage{mathtools}									
	\usepackage{nicefrac}									
	\usepackage{breqn}
	\usepackage[utf8]{inputenc}								
	\usepackage{amsfonts}									
	\usepackage[normalem]{ulem}								
	\usepackage{natbib}
	\usepackage[usenames,dvipsnames,svgnames,table]{xcolor}
	\usepackage[colorlinks]{hyperref}						
	\usepackage{graphicx} 									
	\usepackage{epstopdf} 									
	\usepackage{verbatim}									
	\usepackage{soul}										

\newcommand{\qq}{\langle\bar{q}q\rangle}				
\newcommand{\bp}{\bar{\phi}}
\newcommand{\p}{\phi}
\definecolor{rossoCP3}{cmyk}{0,.88,.77,.40}				

\hypersetup{colorlinks=true, linkcolor=black}

\begin{document}
  	\title{ Safe and free instantons }
  	\author{Francesco~Sannino}
	\email{sannino@cp3.sdu.dk}
	\affiliation{{\color{rossoCP3}CP${}^3$-Origins} \& the Danish Institute for Advanced Study}
 	\author{Vedran~Skrinjar}
	\email{vedran.skrinjar@sissa.it}
	\affiliation{International School for Advanced Studies (SISSA)}
 
\begin{abstract}
We investigate the instanton dynamics of asymptotically safe and free quantum field theories featuring  respectively controllable ultraviolet and infrared fixed points.  We start by briefly reviewing the salient points about the instanton calculus for pure Yang Mills (YM) and  QCD. We then move on to determine the role of instantons within the controllable regime of the QCD conformal window. In this region we add  a  fermion-mass  operator and determine the density of instantons per unit volume as function of the fermion mass.  Finally, for the first time, we extend  the instanton calculus to  asymptotically safe  theories. 
\\
[.3cm]
{\footnotesize  \it Preprint: CP$^3$-Origins-2018-08 DNRF90, 
}
 \end{abstract}

\maketitle

\newpage

\tableofcontents


\hypersetup{colorlinks=true, linkcolor=red}

\section{Introduction}

The standard model and its four dimensional extensions  are described by  gauge-Yukawa theories, it is therefore paramount to understand their dynamics. 

Of special interest are theories that are fundamental according to Wilson~\cite{Wilson:1971bg,Wilson:1971dh}, meaning that {they} are well defined at arbitrarily short distances. Asymptotically free~ \cite{Gross:1973ju,Politzer:1973fx}  and safe \cite{Litim:2014uca}  quantum field theories are two  classes of fundamental quantum field theories. For the former, at extremely short distances, all interactions vanish while for the latter the interactions freeze. In theories with multiple couplings some can be free and others can be safe.  Although asymptotic freedom has a long and successful history, the discovery of four dimensional controllable asymptotically safe quantum field theories is  recent \cite{Litim:2014uca, Litim:2015iea}. This result has enabled novel   dark and bright extensions of the standard model  \cite{Sannino:2014lxa,Abel:2017ujy,Abel:2017rwl,Pelaggi:2017wzr,Mann:2017wzh,Pelaggi:2017abg,Bond:2017wut}. 

The infrared dynamics of fundamental field theories is extremely rich and it can entail confinement and/or chiral symmetry breaking or large distance conformality. This depends on the field content of the specific quantum field theory as well as the presence and type of infrared relevant operators such as scalar and fermion masses. In particular  asymptotically free theories can develop an interacting infrared (IR) fixed point that in certain limits is perturbatively controllable, known as Banks-Zaks (BZ) \cite{Banks:1981nn} fixed point. The  full region in color-flavor space, for gauged fermion theories, where an IR fixed point is present is known as the conformal window, see \cite{Sannino:2009za} for an introduction and \cite{Pica:2017gcb}  for a summary of recent lattice efforts. Recently, building on the large $N_f$ results of \cite{PalanquesMestre:1983zy,Gracey:1996he,Holdom:2010qs,Pica:2010xq,Shrock:2013cca} the concept of conformal window has been extended to include the asymptotically safe region at large number of flavors for which asymptotic freedom is lost \cite{Antipin:2017ebo}. 

The first systematic study of  exact constraints that  a supersymmetric asymptotically safe  quantum field theory must abide including a-maximisation \cite{Intriligator:2003jj} and collider bounds \cite{Hofman:2008ar}  appeared in \cite{Intriligator:2015xxa} extending the results of \cite{Martin:2000cr}.  Here it was also established that Seiberg's SQCD conformal window \cite{Seiberg:1994pq} does not admit an asymptotically safe conformal region. This result is in net contrast with the nonsupersymmetric case  \cite{Antipin:2017ebo}.  Building upon the results of \cite{Intriligator:2015xxa} in  reference \cite{Bajc:2016efj}  the first evidence for supersymmetric  safety was uncovered within the important class of  grand unified theories. The generalisation to  different types of supersymmetric quantum field theories passing all known constraits appeared in   \cite{Bajc:2017xwx}.

 Here we shall be concerned with generalising and applying the instanton calculus to gauge theories in the  perturbative regime of the QCD conformal window as well as of controllable nonsupersymmetric asymptotically safe quantum field theories \cite{Litim:2014uca,Litim:2015iea}.  
 
 To keep the work self-contained we briefly review the instanton calculus for pure Yang Mills (YM) as well as QCD including its large $N_c$ limit in Section \ref{sec:review}. Instantons for the QCD conformal window are introduced and discussed in \ref{sec:BZ}. Here we will consider the two-loop corrected instantons that allow us to follow the perturbative RG flow deep in the infrared where a perturbative interacting IR fixed point occurs. We will then perform our analysis in the fermion-mass deformed theory and derive the main instanton features as function of the fermion mass operator. For example, we shall compute the density of instantons per unit volume as function of the fermion mass measured in units of the RG invariant scale. The latter separates the infrared interacting theory from the UV free fixed point.  Finally we generalise  the instanton calculus to safe rather than free theories  in section \ref{SafeInstantons}. Here we will consider again the fermion mass dependence that now, however, affects the infrared trivial fixed point. We will offer our conclusions in section~\ref{conclusions}. 
  
\section{Instanton calculus review }\label{sec:review}

In quantum field theory (QFT) one aims at computing the partition function,
\begin{equation}\label{part_func_Z[j]}
Z[\mathcal{J}]=\int\mathcal{D}\p \, e^{\mathrm{i}S[\p;\lambda]+\mathcal{J}\p} \ ,
\end{equation}
where $S[\p;\lambda]$ is the sum of a classical action, a gauge-fixing action and a ghost action, depending on the fields $\p$ and the couplings $\lambda$, and $\mathcal{J}$ is a source for $\phi$. If the action is non-integrable one usually attempts to solve the problem through perturbation theory which amounts to expanding the action in powers of small coupling constants $\lambda$. Solutions of the classical theory corresponding to $S[\p;\lambda]$ are specific classical field configurations $\bp$. Since the first variation of the action vanishes on these configurations they represent stationary points, or extrema, of the action. The integrand on the right hand side (RHS) of (\ref{part_func_Z[j]}) is clearly an oscillating function, and thus one may attempt to evaluate the integral by performing an expansion around the classical solution $\bp$. Symbolically, 
we have 

\begin{equation}\label{expand_Z}
Z[\mathcal{J}]=\int\mathcal{D}\p \ e^ {{\mathrm{i}} \left[ S[\bp]+\frac{1}{2}\p S^{(2)}[\bp] \p + \mathcal{O}(\p^3)	\right]    +\mathcal{J}\p} \ . 
\end{equation}
This is the core of the steepest descent method for addressing the issue of oscillating integrals.
One defines the vacuum solution as the  classical configuration that minimizes the energy functional (the Hamiltonian). In the case of (comparatively) simple QFTs there is just one vacuum state and thus there is but a single field configuration $\bp$ around which one should expand the partition function. This is precisely the situation described by equation (\ref{expand_Z}).
\linebreak
For Yang-Mills (YM) theories, often coupled to scalars or fermions, and occasionally coupled to gravity, the vacuum structure is more involved  and if one would naively apply the above prescription several important phenomena would be unaccounted for, such as a deeper understanding of chiral symmetry breaking, the generation of the eta prime mass in QCD, etc. 
 
Let us therefore re-consider briefly the correct approach applicable to a generic QFT~\cite{Shifman:2012zz, Shuryak:1988ck, Coleman1988, Schafer:1996wv, Vainshtein:1981wh}. We begin by Euclideanizing the QFT by performing the Wick rotation { $t\rightarrow\tau=-\mathrm{i}t$}. One should treat gauge fields and fermions with care during this procedure. Euclidean action $S_E$ is a functional of Euclidean fields $\p_E(x)$ living on a 4D Euclidean space described by coordinates $x=(x_1,x_2,x_3,\tau)$. When solving the equations of motion one has to set up the boundary conditions for $|x|\rightarrow\infty$ such that the action remains finite. Usually our conditions require $\p\rightarrow\text{const}$ for $|x|\rightarrow\infty$.
If the potential has only one extremum there is going to be a single vacuum solution (constant field configuration in all of the space) and therefore the naive perturbation theory described by (\ref{expand_Z}) is valid. If, however, the potential has more than one degenerate vacuum, then there exist classical solutions interpolating between these Euclidean vacuua. These finite-action topologically-stable solutions to classical Euclidean equations of motion are called instantons or pseudoparticles \cite{Belavin:1975fg, tHooft:1976snw}. Instantons are topologically stable in the sense that they cannot decay, as going from one such vacuum to another would require bridging an infinite energy barrier \footnote{This is clear because the action is given by the volume integral of the field configuration, so deviations from the instanton configuration automatically pick up infinite contributions.}.

It is now clear that the correct application of the steepest descent method to the Euclideanized version of \eqref{expand_Z} involves a summation over all the instanton configurations. 
 {Even though one does not find instantons as classical solutions to Lorentzian equations of motion, it is clear that the Lorentzian partition function can be obtained by Wick rotating the Euclidean partition function, and thus instantons have to be incorporated in the Lorentzian computation.}
Being  {interpreted as} fields that interpolate between different vacuua, instantons are crucial for understanding a rich vacuum structure in YM theories.
 
\vspace{.5cm}

When discussing instantons   the $SU(2)$ color group plays a special role since $SU(N)$ instantons can be determined starting from the $SU(2)$ case \cite{Bernard:1977nr, Shifman:2012zz}. Let us therefore assume for the moment to have a Euclidean YM action,

\begin{equation}
S[A]=\frac{1}{4}\int_x G_a^{\mu\nu}(A)G_{a\mu\nu}(A)
\end{equation}

where $\int_x\equiv\int d^4x\equiv\int d^3xd\tau$, and $A_\mu^a$ is the gauge field. To find instanton solutions we require the action to be bounded, but rather than asking that $A_\mu^a(x)$ decays faster than $\nicefrac{1}{x}$ for $|x|\rightarrow\infty$, we require it to become a pure gauge,

\begin{equation}\label{inst_asympt_behavior}
A_\mu\xrightarrow{|x| \rightarrow\infty}\mathrm{i}S\partial_\mu S^\dagger \ ,
\end{equation}

where $S$ are $SU(2)$ matrices (not to be confused with the action) that depend on angles only. $SU(2)$ instantons can thereby be seen as maps from $SU(2)$ to itself. Such maps are classified by the third homotopy group and they fall into topologically distinct classes. In the case of $SU(2)$ these are labelled by integer numbers, and members from different classes cannot be continuously mapped into each other \footnote{One can think of a class label as a winding number saying how many times a map winds around the target sphere.}. Instantons belonging to the same class are related by a gauge transformation.

The integers labelling distinct topological classes of instantons can be thought of as topological charges.
Furthermore, for a given instanton configuration the topological charge is given by,

\begin{equation}\label{eq:topological-charge}
n=\frac{g^2}{32\pi^2}\int_x G_a^{\mu\nu}\tilde{G}_{a\mu\nu},\hspace{1cm} n\in\mathbb{Z}
\end{equation}
where $g$ is the gauge coupling. One can complete the square in the action as follows (suppressing indices),

\begin{equation}
S=\frac{1}{4} \int_x GG = \int_x \frac{1}{4} G\tilde{G}+\frac{1}{8}(G-\tilde{G})^2=n\frac{8\pi^2}{g^2}+\frac{1}{8}\int_x(G-\tilde{G})^2
\end{equation}

The action minimum for the instanton of topological charge $n$ clearly corresponds to the value\footnote{Strictly speaking equation \eqref{eq:finite-action-value} holds for positive $n$. Negative values of  $n$ are obtained via a parity transformation, since then $G\tilde{G}\rightarrow-G\tilde{G}$. Following the same argument as above the action attains its minimum at $|n| \frac{8\pi^2}{g^2}$ for the field configuration which is anti self-dual, $G=-\tilde{G}$. Such field configuration is called anti-instanton.}

\begin{equation}\label{eq:finite-action-value}
S|_{\text{n-instanton}}=n \frac{8\pi^2}{g^2}.
\end{equation}

This is achieved when the field satisfies the self-duality condition, $G=\tilde{G}$. Using the Bianchi identities, one can show that the field satisfying the self-duality condition is on-shell, i.e. it automatically satisfies the equations of motion. Computing the value of the action on an instanton solution constitutes the first important result of the instanton calculus.
Starting from the asymptotics, (\ref{inst_asympt_behavior}), and assuming the same directional dependence of the solution in all spacetime points one can write an ansatz for the instanton. Requiring absence of singularities at the origin of space and self-duality of the solution suffices to uniquely fix the instanton (up to collective coordinates) \cite{Shifman:2012zz}.  This is the famous BPST instanton ($SU(2)$ instanton with charge n=1) \cite{Belavin:1975fg}. Explicitly,

\begin{equation}\label{BPST solution}
A_\mu^a=\frac{2}{g}\eta_{a\mu\nu}\frac{(x-x_0)^\nu}{(x-x_0)^2+\rho^2} \ .
\end{equation}
The above expression for the BPST instanton is in the so-called regular gauge. The parameter $\rho$, the instanton size, is the aforementioned integration constant and it is one of the instanton collective coordinates. The remaining collective coordinates are instanton position in spacetime, $x_0$, and its orientation in color space. Finally, $\eta_{a\mu\nu}$ are known as 't Hooft symbols \cite{tHooft:1976snw}.

The generalisation to simple Lie algebras is obtained directly from the $SU(2)$ BPST instanton exploiting the fact that any $SU(N)$ group contains $SU(2)$ subgroups. To deduce the $SU(N)$ instantons one simply embeds the BPST solution \eqref{BPST solution} into $SU(N)$. This choice of embedding is ambiguous, but the most common choice is the so-called minimal embedding. It consists in taking the $SU(N)$ generators in the fundamental group, and taking the first three generators $T^1,...,T^3$ to be block-diagonal with $SU(2)$ generators embedded in the upper-left corner. The $SU(N)$ BPST instanton is obtained by contracting the first three generators $T^a$, $a=1,2,3$ with the BPST solution (\ref{BPST solution}). One can analogously obtain $SU(N)$ instantons with charge $n\neq 1$ from other $SU(2)$ solutions. This simple prescription works because the third homotopy group of $SU(N)$ is $\mathbb{Z}$ for all $N$, and with the minimal embedding each equivalence class of $SU(N)$ solutions contains a representative $SU(2)$ instanton.

\subsection{The QCD story}
Here we shall see that an instanton ensemble plays an important role in determining  the structure of the QCD vacuum.  We will start with reviewing the construction of a partition function for such an ensemble. We begin with the famous result for the one instanton partition function which was given by 't Hooft in 1976 \cite{tHooft:1976snw}.

The vacuum-to-vacuum transition amplitude in presence of a single instanton is given by the following 1-loop instanton calculus result for an $SU(N_c)$ pure Yang-Mills theory 
\cite{tHooft:1976snw, Bernard:1979qt}, 

\begin{eqnarray}
\label{eq:one-instanton-density}
	W^{(1)}&=&\frac{4}{\pi^2} 
			\frac{\exp\left(-\alpha(1)-2(N_c-2)\alpha(\nicefrac{1}{2})\right)}{(N_c-1)!(N_c-2)!}
			\int d^4x d\rho \rho^{-5} \left(\frac{4\pi^2}{g_0^2}\right)^{2N_c}	\exp \left(-\frac{8\pi^2}{g_{1L}^2}\right)			\\
\label{eq:one-instanton-density-compact}
			&\equiv &C_c \int d^4x d\rho \rho^{-5} \left(\frac{8\pi^2}{g_0^2}\right)^{2N_c}	\exp \left(-\frac{8\pi^2}{g_{1L}^2}\right)
\end{eqnarray}
The integral on the RHS is over the instanton size $\rho$, and its integrand is refferred to as the instanton density. {Note that the numerical factor $C_c$ depends only on the number of colors and it also contains the factor $2^{-2N_c}$.} 
 The above integral is IR divergent ($\rho\rightarrow\infty$,  {see (\ref{eq:mastereq-M})}) because of the running coupling in the exponent. 
Clearly one has to tame this behavior for the result to be meaningful.  

If the Yang-Mills theory is coupled to $N_f$ Dirac fermions then, at one loop, they contribute via the fermion determinant to the above result. It is both possible and useful to separate the zero and non-zero fermionic modes. The non-zero modes contribute to the exponential as \cite{tHooft:1976snw},

\begin{equation}\label{eq:fermion-nonzero-modes}
\exp\left[-\frac{2N_f}{3}log(\rho / \rho_0)+2N_f\alpha(1/2) \right] \ ,
\end{equation}
where the first term is the fermion contribution to the 1-loop running of the gauge coupling and $\alpha(x)$ is a function defined in \cite{tHooft:1976snw}\footnote{$\alpha(1/2)=0.145873$ and $\alpha(1)=0.443307$}. Taking all the fermions to have the same mass $m$, the zero modes contribute a term

\begin{equation}
\label{eq:ZeroModes}
(m \rho)^{N_f}.
\end{equation}
We can now generalise the result in \eqref{eq:one-instanton-density-compact} to include fermions using the 1-loop running of the QCD gauge coupling, 
\begin{equation} \label{eq:1Lbeta}
\frac{8\pi^2}{g_{1L}^2}=\frac{8\pi^2}{g_0^2}-b \ log(\rho/\rho_0) \ ,  \quad {\rm with  } \quad b=\frac{11}{3}N_c-\frac{2}{3}N_f \ ,
\end{equation}
and derive
\begin{eqnarray} \label{eq:mastereq-M}
W^{(1L)}&=& \frac{4}{\pi^2}\frac{\exp(-\alpha(1)+4\alpha(\nicefrac{1}{2}))}{(N_c-1)!(N_c-2)!}
			\exp(2(N_f-N_c)\alpha(\nicefrac{1}{2}))\times \nonumber								\\
	&\times & m^{N_f}(\frac{4\pi^2}{g_0^2})^{2N_c}\int d^4x d\rho \rho^{-5+N_f}
			\exp(-\frac{8\pi^2}{g_0^2}+(\frac{11}{3}N_c-\frac{2}{3}N_f)log(\rho/\rho_0))		\\
				\label{eq:mastereq-M-1}
	&=&C_{cf}\ m^{N_f}\int d^4x d\rho \rho^{-5+N_f}(\frac{8\pi^2}{g_0^2})^{2N_c}
			\exp(-\frac{8\pi^2}{g_{1L}^2}) \ .
\end{eqnarray}

Besides {suffering from} the divergence of the instanton density for large instantons, the master equation (\ref{eq:mastereq-M})  {has another important feature}.
The zero mode contributions  of (\ref{eq:ZeroModes}) imply the vanishing of the whole amplitude as $m\rightarrow 0$. This was noted and thoroughly discussed in \cite{tHooft:1976snw}, see also \cite{Shifman:1979uw, Diakonov:1985eg}.

A simple strategy to bypass this problem was initially given by \cite{Shifman:1979uw}. Their reasoning goes as follows. It is empirically known that the QCD vacuum is a medium in which many condensates form, so instead of studying a single instanton in isolation one should take into account the condensation phenomenon on the instanton density. In particular, the authors focussed on the chiral condensate $\qq$. At the time one could not determine the chiral condensate from first principles, so the authors employed its phenomenological value. Besides its relevance as an order parameter for the spontaneous breakdown of chiral symmetry (SBCS), one  considers the chiral condensate as a dynamical fermion mass that should be used in the amplitude \eqref{eq:mastereq-M} instead of the bare mass. Following \cite{Shifman:1979uw} we compute the effective quark mass in presence of a non-vanishing $\qq$ condensate in QCD to be given by,

\begin{eqnarray}
m_{eff}	&=&m-\frac{4\pi^2\rho^2}{N_c}\langle\bar{q}^Lq^R\rangle	\\
		\label{eq:Meffective}
		&=&m-\frac{2\pi^2\rho^2}{N_c}\langle\bar{q}	(1+\gamma_5)	q\rangle
\end{eqnarray}

 Crucially, in the case $m\rightarrow0$ the effective mass doesn't vanish, meaning that the vacuum-to-vacuum transition amplitude in the presence of a single instanton is non-zero provided that the chiral condensate forms. In this way  the authors of \cite{Shifman:1979uw} successfully pointed towards the physical mechanism responsible for resolving the issue with the zero mass limit.

\vspace{.25cm}

Let us now return to the other issue, the IR divergences of the instanton density. Conceptually, it is reasonable to expect that if QCD forms gluon condensates, then they should be described by a statistical ensemble of the instantons forming them. The early attempts in this direction imagined the QCD vacuum to be described by an instanton gas \cite{Callan:1977gz, Vainshtein:1981wh}. This was demonstrated to be a poor description of the physical vacuum, since instantons were much more strongly interacting. The solution came in the form of Shuryak's instanton liquid model in 1982 \cite{Shuryak:1982dp}. He had shown that a simple model of the instanton medium as a liquid with only two free parameters can effectively explain a number of nuclear physics observables. His model assumes that all instantons have the same size, $\bar{\rho}$, and he  obtained the instanton size and the density of the instanton liquid from the empirical value of the gluon condensate. The approach thus doesn't explain why the instanton density is a delta-like peak around some $\bar{\rho}$, but such a description has predictive power and seems to explain nuclear physics data well.

The above ideas were developed more systematically within the mean field approximation by Diakonov and Petrov in 1983 and 1985,  aiming at a description of an ensemble of instantons from  first principles. Failure of the instanton gas picture had implied that the instanton interactions should be modelled even if the medium itself will turn out to be rather dilute. This was motivated by the expectation that the instanton interactions would remove the IR divergence. Because of the above, the authors in \citep{Diakonov:1983hh} introduced a modified variational procedure in an attempt to approximate the exact multi-instanton partition function. They applied their method to pure Yang-Mills theory and besides curing the IR problem they also successfully computed a number of physical observables. In a later work, \cite{Diakonov:1985eg}, the method was extended to include gauged fermions. The central result of this paper is that in an instanton background fermions develop a momentum dependent effective mass which is non-vanishing in the zero momentum limit confirming the expectations of reference \cite{Shifman:1979uw} as summarised above.

\vspace{.25cm}

Before moving on to the large-$N_c$ theory  we will comment on one more issue regarding the master equation (\ref{eq:mastereq-M-1}). It follows from a 1-loop computation that the coupling in the exponential term is renormalized, but the one in the pre-exponential factor is not.  In literature, this problem is  often addressed by recognizing that at two loops the pre-exponential factor gets renormalized \cite{Diakonov:1983hh} and thus one replaces the bare coupling by the 1-loop running coupling, and the 1-loop coupling by the 2-loop coupling. For completeness we also provide the standard result for the two loop running coupling 
\cite{Caswell:1974cj, Shifman:1979uw},

\begin{eqnarray}\label{eq:naive-2-loop-running}
\frac{8\pi^2}{g_{2L}^2}&=&\frac{8\pi^2}{g_0^2}- b \log{\rho/\rho_0}+\frac{b'}{b}\log(1-\frac{g_0^2}{8\pi^2}\log\rho/\rho_0) \ ,		\\
b'&=&\frac{51}{9}N_c^2-\frac{19}{3}N_f \ .	 
\end{eqnarray}

Note that the behavior of the coupling given in (\ref{eq:naive-2-loop-running}) is not the exact 2-loop one. In fact, this is only the leading UV contribution valid in the deep UV regime for the asymptotically free phase of QCD \footnote{This is clear since the expression (\ref{eq:naive-2-loop-running}) is manifestly ignorant of the possible existence of a perturbative IR fixed point.}. We will elaborate more on this point in section \ref{sec:BZ}.

\subsection{Large $N_c$ }\label{sec:pure-YM}
Pure Yang-Mills theory at large-$N_c$ is an important step towards studying instantons in the conformal window as well as asymptotically safe instantons. In fact, many of the formulae derived in this subsection can be adapted to include the effects of fermions in these theories. Herein we briefly outline the variational approach of \cite{Diakonov:1983hh} and present their main results. We particularly focus on the large-$N_c$ limit following reference \cite{Schafer:2002af}.

\vspace{.5cm}

Assuming that the pure YM vacuum is given by a background gauge field configuration which consists of a large set of instantons, following  \cite{Diakonov:1983hh}, in absence of exact results in pure YM theory,   one approximates such a background to be a sum of simple, localized 1-instanton solutions.  Starting from such an ansatz  the ground state can be derived by introducing a modification of the Feynman's variational principle. This consists in taking an action S, modifying it slightly to get an action $S_1$ s.t. it has a minimum on our ansatz field configuration, and then using the fact that
\begin{equation}\label{eq:variational_principle}
Z\geq Z_1 e^{-\langle S-S_1 \rangle} \ ,
\end{equation}
\noindent
where Z is the partition function that we want to approximate using the variational principle. Z is given by

\begin{equation}
Z=\int \mathrm{D}\phi e^{-S[\phi]}\ ,
\end{equation}
and $Z_1$ is defined analogously, with the action $S_1$. The expectation values $\langle\ .\ \rangle$ are taken with respect to  the measure $\exp(-S_1)$.

Let us take the background field to be given by $\bar{A}=\sum_I A_I + \sum_{\bar{I}}  A_{\bar{I}}$  where I runs over the instanton configurations and $\bar{I}$  over anti-instantons. We may rewrite the Lagrangian as follows,

\begin{eqnarray}
-\frac{1}{4g^2}F^2(\bar{A})&=&-\frac{1}{4g^2}\left(\sum_{i=I,\bar{I}}F^2(A_i)+ F^2(\bar{A})-\sum_{i=I,\bar{I}}F^2(A_i)\right)	\\				
							& \equiv & -\frac{1}{4g^2}\left(\sum_{i=I,\bar{I}}F^2(A_i)+U_{int}\right) \ ,
\end{eqnarray}
where the first term is the Lagrangian of a non-interacting instanton gas, and the second term describes the interaction in the medium. 
 {From here on we use notation} $\nicefrac{1}{4g^2}F^2=\nicefrac{1}{4}G^2$. 
Including the bosonic statistics factors $N_{\pm}$ in front of the partition function, normalizing both sides of (\ref{eq:variational_principle}) to the perturbation theory vacuum, and regularizing the determinants, at one loop order we obtain the following expression,

\begin{eqnarray}
\label{eq:RILM-partition-func-1}
\left. \frac{Z}{Z_{ptb}}\right|_{reg, 1L}	&\geq &	\frac{1}{N_+!N_-!}\int\prod_i^{N_++N_-}d\gamma_i \ d(\rho_i) e^{-\beta(\bar{\rho})U_{int}(\gamma_i)}	\\
\label{eq:RILM-partition-func-2} 
							&\equiv & \frac{1}{N_+!N_-!}\int\prod_i^{N_++N_-}d\gamma_i \ e^{-E(\gamma_i)} \ . 
\end{eqnarray}
In this expression $\gamma_i$ represents the collective coordinates of the i-th pseudoparticle (see  \ref{sec:review}). $d(\rho)$ stands for the 1-instanton density (\ref{eq:one-instanton-density-compact}), and we use the standard notation, 

 \begin{equation}
\beta(\rho)\equiv8\pi^2/g^2(\rho)\ .
\end{equation}
In the expression (\ref{eq:RILM-partition-func-1}) $\beta(\rho)$ is renormalized by 1-loop determinants at a scale $\bar{\rho}$ corresponding to the average instanton size. In the second line, (\ref{eq:RILM-partition-func-2}), we've introduced the compact notation,

\begin{equation}\label{RILM-energy}
E(\gamma_i)=\beta(\bar{\rho})U_{int}(\gamma_i)-\sum_i \log d(\rho_i) \ .
\end{equation}
If the medium is sufficiently dilute one can consider only two-particle interactions in the interaction term, all the other ones being subdominant \footnote{In fact first corrections to this computation come not from considering higher order interactions but from considering 2-loop beta functions \cite{Diakonov:1983hh}.}. This is the key physical ingredient beyond the simple instanton gas model. The interaction potential has been determined in \cite{Diakonov:1983hh}. Integrating over the relative angle between two instantons in color space, and integrating over the instanton separation one obtains a remarkably simple expression,

\begin{eqnarray}
U_{int}^{2-body}(\rho_1,\rho_2)=\gamma^2 \rho_1^2 \rho_2^2\ ,\qquad \gamma^2=\frac{27\pi^2}{4}\frac{N_c}{N_c^2-1}\ ,
\end{eqnarray}
where $\rho_{1,2}$ are the sizes of the two pseudoparticles, and the coupling $\gamma^2$ has the characteristic $1/N_c$ behavior.

We may now use the variational principle. Assuming that the effect of the 2-body interactions can be well captured by a modification of the 1-instanton densities $d(\rho)$, we write,

\begin{equation}
E_1(\gamma_i)=-\sum_I^{N_+} \log\mu_+(\rho_I)-\sum_{\bar{I}}^{N_-} \log\mu_-(\rho_{\bar{I}}) \ .
\end{equation}

\noindent
Substituting $E_1$ in place of $E$ in (\ref{eq:RILM-partition-func-2}) we get,

\begin{equation}\label{eq:RILM-partition-function-3}
Z_1=\frac{1}{N_+!N_-!}V^{N_++N_-}(\mu_+^0)^{N_+}(\mu_-^0)^{N_-} \ ,
\end{equation}

\noindent
where,

\begin{equation}
\mu_\pm^0=\int_0^\infty d\rho \  \mu_\pm(\rho) \ .
\end{equation}
To apply the variational principle to find the optimal value of $\mu(\rho)$ we start by evaluating $\langle E-E_1\rangle$ which enters in (\ref{eq:variational_principle}). First we express $\langle E-E_1\rangle$  in terms of,

\begin{equation}\label{eq:rho-square-bar}
\overline{\rho_{\pm}^2}=\frac{1}{\mu_{\pm}^0}\int d\rho \ \rho^2 \mu_{\pm}(\rho) \ .
\end{equation}
Next, we minimize (\ref{eq:RILM-partition-function-3}) wrt $\mu_\pm$. There's an arbitrary constant appearing in the optimal $\mu_\pm$, and if these are chosen equal then $\mu_+=\mu_-\equiv\mu$. Writing $N_++N_-=N$, we find the optimal $\mu$ to be,

\begin{equation}\label{eq:naive-optimal-mu}
\mu(\rho)=d(\rho)\exp\left(-\frac{\beta \gamma^2 N}{V}\overline{\rho^2}\rho^2\right),
\end{equation}

\noindent
where $\beta\equiv\beta(\bar{\rho})=8\pi^2/g^2(\bar{\rho})$. This can be reinserted in (\ref{eq:rho-square-bar}) to give,

\begin{equation}\label{eq:rho-square-bar-1}
(\overline{\rho^2})^2=\frac{\nu}{\beta \gamma^2 N/V}, \ \ \ \ \ \nu=\frac{b-4}{2}.
\end{equation}

This expression can be further inserted in the optimal $\mu$, and $\mu^0$ can be easily found using the explicit form of the optimal $\mu$ and of the 1-instanton density. Finally we can determine the RHS of \eqref{eq:RILM-partition-function-3}, see \cite{Diakonov:1983hh} for more details.

Instead of keeping the number of pseudoparticles $N$ fixed we can work in the grand canonical ensemble. This allows us to find the average number of instantons in the medium by maximizing the RHS of (\ref{eq:RILM-partition-function-3})  as a function of $N$. For the bosonic factors we set $N_\pm!=(N/2)!$ and use the Stirling approximation. This brings us to the following important expression for the average instanton number,

\begin{eqnarray}\label{eq:avg-instanton-number}
\langle N \rangle = V \Lambda_{YM}^4 \left(\Gamma(\nu) C_{cf} \tilde{\beta}^{2N_c} (\beta \gamma^2 \nu)^{-\nu/2}\right)^{\frac{2}{\nu+2}}\ ,
\end{eqnarray}
where $\tilde{\beta}=8\pi^2/g_0^2$. Note that $\overline{\rho^2}$ enters this equation through $\beta=8\pi^2/g^2_{1L}(\overline{\rho^2})$, so (\ref{eq:rho-square-bar-1}) and (\ref{eq:avg-instanton-number}) should be solved simultaneously (consistently). 
The importance of the average number of instantons comes from the fact that it is related both to the gluon condensate, vacuum energy, topological susceptiblity, and in a theory with fermions, to the $U(1)$ axial anomaly.

\begin{figure}
\includegraphics[scale=.7]{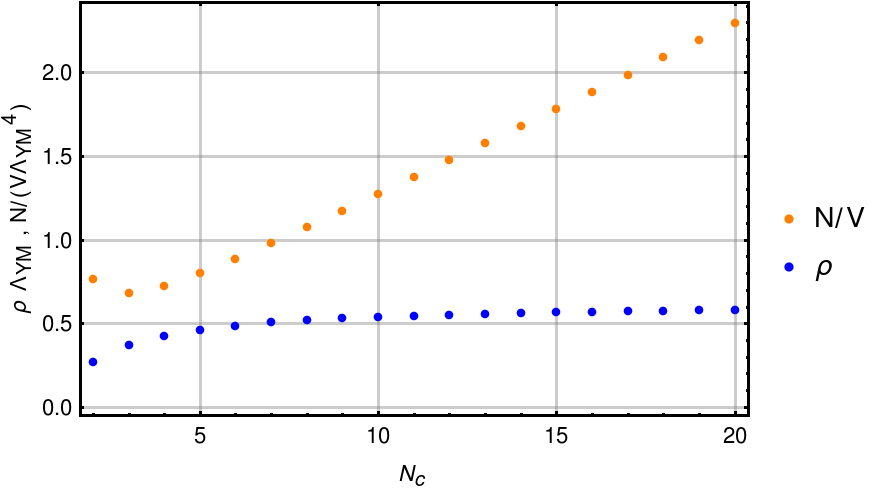}
\caption{\label{fig:rho-and-density}
Instanton size and density of instantons as function of $N_c$}
\end{figure}

In terms of the number of instantons per unit volume the partition function takes the following simple form,

\begin{equation}\label{eq:RILM-partition-function}
Z=\exp \left[ \frac{1}{2} (\nu+1) \langle N \rangle \right].
\end{equation}
We can solve numerically for the expectation values of the instanton size and of the density of instantons in the vacuum. To do that we need to perform the aforementioned RG improvement by promoting $\tilde{\beta}$ to $\beta$ and $\beta$ to $8\pi^2/g^2_{(2L)}(\bar{\rho})$. 
{Note that it is useful to introduce a} 
free parameter $a$, called fudge factor, {in the log term of the 1-loop running coupling (\ref{eq:1Lbeta}). The fudge factor}  essentially parametrizes the uncertainty on the actual confining scale $\Lambda_{YM}$. The numerical results are shown in Figure \ref{fig:rho-and-density}. Even for modest values of $N_c$ shown in the figure one already notices that the density of instantons increases as $\mathcal{O}(N_c)$, whereas the average instanton size is quite independent of $N_c$ and is always of $\mathcal{O}(1)$.

\begin{figure}
\includegraphics[scale=.7]{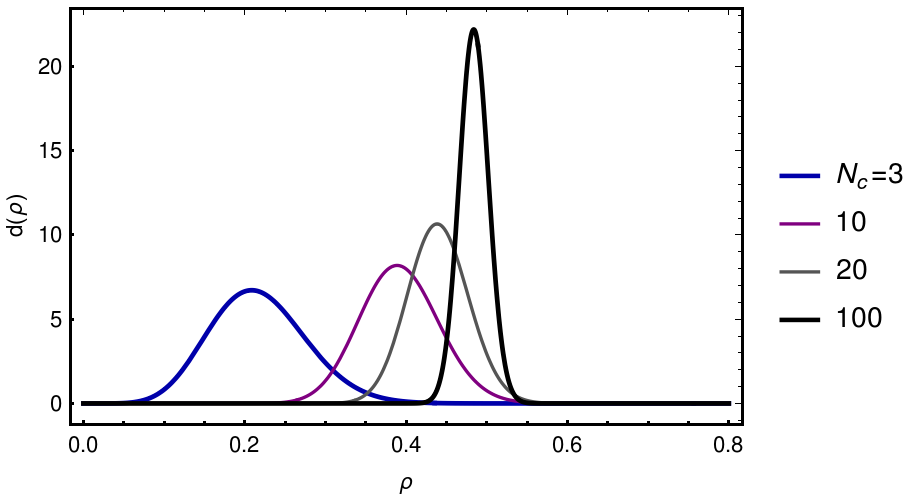}
\caption{\label{fig:instanton-density}
Effective instanton density profile as a function of $N_c$. (Normalized to unity.)}
\end{figure}

We can also study the dependence of the effective instanton density $d(\rho)$ on the number of colors $N_c$. The results are shown in Figure \ref{fig:instanton-density}. Already from (\ref{eq:one-instanton-density}) we know that the amplitude decreases rapidly with $N_c$, but what we consider here is the shape and the spread of the distribution. (To this end we normalize all the distributions to $\mu^0=1$.) In particular, we notice that the distribution has a prominent peak centered about the average instanton size, and in the large $N_c$ limit becomes essentially delta like \cite{Schafer:2002af}.

Recall the relation between the full Lagrangian and the instanton gas Lagrangian, $F^2=\sum_i F_i^2+32\pi^2 U_{int}$. Since we know the value of the action for a BPST instanton (see eq. (\ref{eq:finite-action-value})), we know that

\begin{equation}\label{eq:RILM-energy}
\langle \int \frac{d^4x}{32\pi^2} F^2 \rangle = \langle N \rangle + \langle U_{int} \rangle \ ,
\end{equation}

\noindent
and from (\ref{eq:RILM-partition-func-1}) it follows $\langle U_{int} \rangle = - \partial \log Z / \partial \beta$. From (\ref{eq:RILM-partition-function}) we thus obtain,

\begin{equation}
\langle U_{int} \rangle = \frac{\nu}{2\beta}\langle N \rangle\ .
\end{equation}

\begin{figure}
\includegraphics[scale=.7]{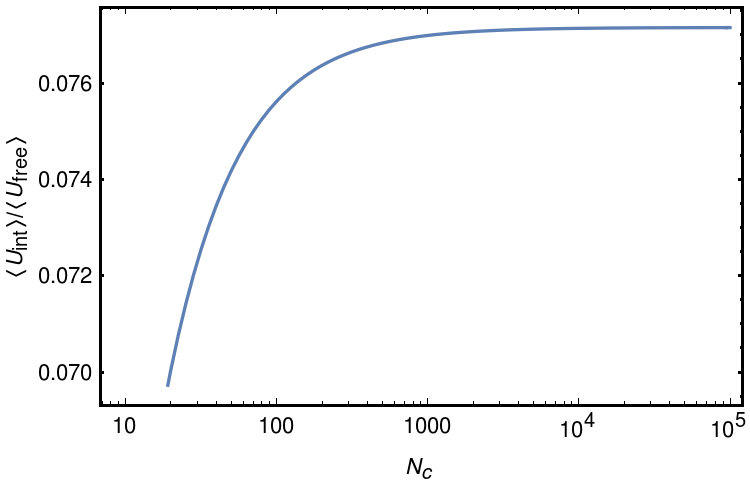}
\caption{\label{fig:YM-energy-ratio}
Ratio of interaction energy to free energy as a function of $N_c$. We've fixed a=1/10.}
\end{figure}
\noindent
Figure \ref{fig:YM-energy-ratio} shows the ratio of the interaction energy to free energy. Since the free energy is larger than the interaction energy we  can trust the simplified 2-body interaction model.
Further, because the gluon field VEV is related to the trace of the stress energy tensor (SET) by the scale anomaly relation, and since the trace of the SET is in a direct relation to the vacuum energy density, we obtain the following leading-order expression for the vacuum energy density,

\begin{equation}\label{eq:RILM-vacuum-energy}
\mathcal{E}=-\frac{b}{4}\frac{\langle N \rangle}{V} \ .
\end{equation}
Notice that it grows quadratically with $N_c$, with an additional factor of $N_c$ with respect to non-interacting instanton gas \cite{Schafer:2002af}.

Let us now compute the topological susceptibility. This is of particular interest because it is an observable.
We start by adding the topological theta-term, $\frac{i \theta}{32\pi^2} \int d^4x F \tilde{F}$, to the action. The topological susceptibility is defined by,

\begin{equation}\label{eq:RILM-top-suscept}
\chi_{top}=-\frac{\partial^2 \log Z}{\partial \theta ^2}|_{\theta =0} =\langle \left(  \int d^4x \frac{F \tilde{F}}{32\pi^2} \right)^2 \rangle \ .
\end{equation}

\noindent
In particular, adding the $\theta$-term to the partition function doesn't modify the computation of $\mu(\rho)$, or $\overline{\rho^2}$, and thus the only modification to (\ref{eq:RILM-partition-function}) is in an additional term $+i\theta (N_+-N_-)$. Self-consistently, by rewriting this as 
\begin{equation}
Z=\exp \left[ \frac{\nu+2}{2} \langle N \rangle (1-\frac{\theta^2}{\nu+2}+\mathcal{O}(\theta^4)) \right] \ ,
\end{equation}
and taking the derivative as in (\ref{eq:RILM-top-suscept}) we get \cite{Diakonov:1983hh},
\begin{equation}\label{eq:RILM-top-suscept-result}
\chi_{top}=\langle N \rangle \ .
\end{equation}

We are now ready to investigate and extend the role of instantons within the conformal window of QCD and for asymptotically safe quantum field theories.

\section{Conformal Window Instantons}\label{sec:BZ}

In this section we determine the instanton dynamics in the QCD IR conformal window.  We shall be prevalenty concerned with the calculable part of the conformal window, the one in which an IR fixed point is reached perturbatively and that is often referred to as a l\'a Banks-Zaks  \cite{Banks:1981nn}.
  
\begin{figure}
\centering
\includegraphics[scale=.7]{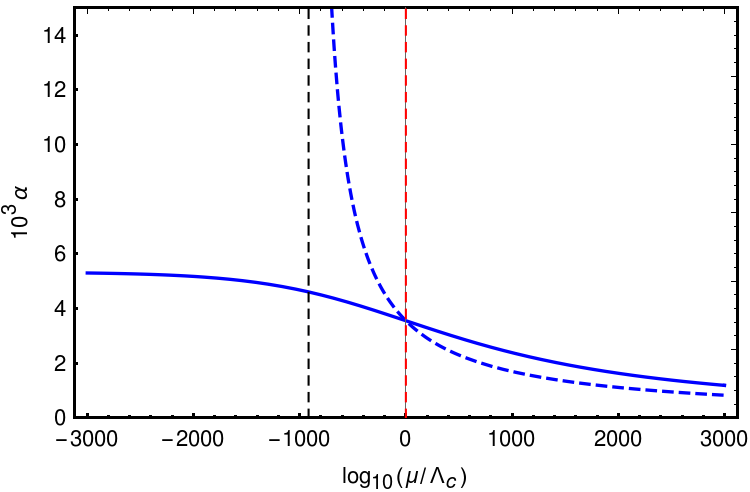}
\caption{\label{fig:BZvsQCD-running}
Banks-Zaks running, shown in continuous blue line for $\epsilon=-1/10$, interpolates between an interacting fixed point in the IR and a non-interacting fixed point in the UV. This result is obtained starting at two loops, whereas the analogous one loop running is given in blue dashed line. The only scale in the BZ theory is the RG-invariant scale $\Lambda_c$ corresponding to the red dashed line. We chose the matching conditions s.t. the one loop and the two loop couplings match at the scale $\Lambda_c$; this fixes the one loop IR divergence scale shown in black dashed line.  }
\end{figure}

The perturbative IR fixed point occurs for the number of fermions $N_f$ tuned to be slightly below $\nicefrac{11}{2}N_c$ in the large-$N_c$, large-$N_f$ limit.  In this limit  one introduces an expansion in the physical parameter  $\epsilon$, defined in (\ref{eq:def-epsilon}), that measures the distance, in flavor space, from the loss of asymptotic freedom. This parameter can be made arbitrarily small. The fixed point value, being an expansion in $\epsilon$ can be made arbitrarily weakly interacting rendering the expansion controllable. In figure  \ref{fig:BZvsQCD-running} we compare the running coupling in the Banks-Zaks theory for $\epsilon=-1/10$ at one loop (diverging) and at two loops (converging to a fixed point).

It is immediately clear from the running of the coupling that the infrared dynamics, being conformal, is quite distinct from the chiral symmetry breaking QCD scenario. 
 In particular, in the IR instead of becoming non-perturbatidve it remains within the perturbative regime until it finally reaches a conformal theory in the deep infrared. We shall consider the epsilon regime in which the two-loops analysis remains trustworthy. It would be interesting to extend the present work to  higher loops 
 \cite{Pica:2010xq,Ryttov:2010iz,Ryttov:2016ner}.   Due to the perturbative control, we can fully include the fermion effects at one loop order by including their contribution to the beta function of the gauge coupling. 
 
\begin{figure}
\centering
\includegraphics[scale=.7]{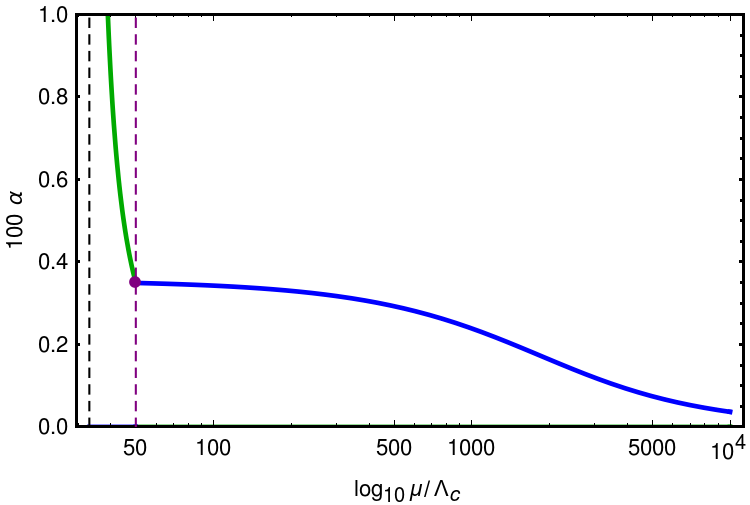} \\
\caption{\label{fig:massive-BZ-running}
Blue line shows the Banks-Zaks running for $\epsilon=-1/10$, and green line corresponds to the pure YM running. Purple dot shows the matching couplings at the fermion mass scale which is given by purple dashed line. Black dashed line is the scale $\Lambda_{YM}$, and scale $\mu=\Lambda_c=1$ can't be shown due to the use of log scale on the horizontal axis.}
\end{figure}

 It is particularly interesting to investigate the mass-deformed perturbative conformal field theory as argued first in \cite{Sannino:2010ca}.    Starting with  fermions, all of the same mass $m\ll\Lambda_c$, the running is given in the top panel of figure \ref{fig:massive-BZ-running}. In the deep UV, at energies higher than the fermion mass $m$, the running is dominated by the free fixed point.  At energies below the fermion mass fermions can be integrated out.  In the perturbative regime of the conformal window we can follow the perturbative flow down to $m$.  At energies lower than the common fermion mass one enters the YM regime. In a mass-independent scheme (although our results for physical quantities are scheme independent) one matches the pure YM coupling with the one with massless fermions at the scale $m$, and the YM running takes over as shown in the bottom panel of figure \ref{fig:massive-BZ-running}.

In QCD one needs to take particular care of the low-energy fermion modes  when the hard common fermion mass is sufficiently small. This is so since these modes are delocalized and feel the presence of the instanton medium. In the perturbative regime of the conformal window one can continue lowering the fermion mass all the way to zero because the coupling is guaranteed to stay perturbative down to the fermion mass scale. Above the common fermion mass energy  no condensate can form because the theory can be made arbitrarily weakly interacting \cite{Diakonov:1985eg}. As the fermions become massless  we expect the instantons to {\it melt away} and the vacuum-to-vacuum transition amplitude due to instantons to vanish.

\vspace{.5cm}
To take into account the full perturbative running we consider the RG-improved master equation. The often used naive 2-loop running  approach (\ref{eq:naive-2-loop-running}) is valid only in the deep UV since it does not account for the Banks-Zaks  IR fixed point. 
Let us thus look into the exact 2-loop RG running more closely.

We begin by defining the 't Hooft coupling,

\begin{equation}\label{eq:tHooft-coupling}
\alpha=\frac{g^2N_c}{(4\pi)^2}.
\end{equation}

\noindent
The two loop beta function of the gauge coupling in presence of fermions can be written as,

\begin{equation}\label{eq:2-loop-betas}
\mu \partial_\mu \alpha\equiv\beta_\alpha=-B \alpha^2 + C \alpha^3.
\end{equation}

\noindent
Here $B=-\nicefrac{4}{3}\ \epsilon $ and $C=25+\nicefrac{26}{3}\ \epsilon$, and the physical control parameter is given by 
\begin{equation}\label{eq:def-epsilon}
\epsilon=\frac{N_f}{N_c}-\frac{11}{2}<0 \ .
\end{equation}

\noindent
The exact 2-loop running is given by  \cite{Litim:2015iea} 

\begin{equation}\label{eq:2-loop-running}
\alpha(\mu)=\frac{\alpha_*}{1+W(z(\mu))}\ ,
\end{equation}

\noindent
where 

\begin{equation}
\alpha_*=B/C
\end{equation} 

\noindent
is the IR Banks-Zaks fixed point, and W stands for the Lambert (or productlog) function. z($\mu$) will be defined shortly. Expansion around $\mu\rightarrow\infty$ yields equation (\ref{eq:naive-2-loop-running}).
The running stemming from \eqref{eq:2-loop-running} is manifestly  bounded and it interpolates between $\alpha=0$ for infinite energies and $\alpha=\alpha_*$ in the IR, as it can be seen from figure \ref{fig:BZvsQCD-running}.

Let us note that $\partial_\alpha\beta_\alpha(\alpha)$ vanishes for $\alpha=\nicefrac{2}{3}\ \alpha_*\equiv\alpha_c$. The scale at which one reaches this value of the coupling is critical in the sense that at this scale the gauge coupling changes scaling from canonical to a non-Gaussian one. This scale,

\begin{equation}\label{eq:2-loop-RG-inv-scale}
\mu(\alpha_c)\equiv\Lambda_c=(2e^{-\frac{1}{2}})^{-1/ \theta_*}	 (1-\frac{\alpha}{\alpha_*})^{-1/\theta_*}\ \mu \ ,
\end{equation}

\noindent
is the 2-loop RG-invariant scale in the sense that $\mu\partial_\mu(\Lambda_c)=0$ (to linear order), and $\theta_*$ is the {eigenvalue of the RG flow} at the  FP,
\begin{equation}
\theta_*=\frac{\partial \beta_\alpha}{\partial \alpha}|_{\alpha_*}=\alpha_* B \ .
\end{equation}

Inserting the RG-improvements in the 1-loop master equation (\ref{eq:mastereq-M-1}) we have,

\begin{eqnarray}
d_{2L}(\rho)&=&C_{cf} m^{N_f} \rho^{N_f-5}(b \log M\rho)^{2N_c} e^{-\frac{8\pi^2}{g^2_{2L}}}		\\
		&=& C_{cf}   \exp(1/2-\log 2)^{-\frac{8\pi^2}{g^2_*}}m^{N_f}\rho^{N_f-5}(\rho \Lambda_c)^{\frac{1}{2}BN_c}\times	\nonumber \\
		&\times &	(b \log M\rho)^{2N_c} W(z(\rho))^{\frac{8\pi^2}{g_*^2}}.	\label{eq:2-loop-inst-density} 
\end{eqnarray}

\noindent
In the above expression we used $C_{cf}$ as defined in (\ref{eq:mastereq-M}), $\Lambda_c$ defined in (\ref{eq:2-loop-RG-inv-scale}),

\begin{equation}\label{eq:arg-of-Lambert}
z(\rho)=e^{1/2-\log 2}(\rho\Lambda_c)^{-\alpha_*B},
\end{equation}

\noindent
1-loop beta coefficient $b$ given in \eqref{eq:1Lbeta}, and 1-loop RG-invariant scale M,

\begin{equation}\label{eq:1-loop-RG-inv-scale}
M=\frac{1}{\rho_c}\exp{(-\frac{1}{b}\frac{8\pi^2}{g(\rho_c)^2})}=\Lambda_c \exp{(-\frac{3}{2}\frac{C}{B^2})}.
\end{equation}

Setting $\rho^2\rightarrow\overline{\rho^2}$ in second line of (\ref{eq:2-loop-inst-density}) the expression for the instanton density takes a form which is similar to what we had in the pure YM case.
 In fact, defining,

\begin{equation}\label{eq:f-of-rho-bar}
f(\overline{\rho})=C_{cf}  \exp(1/2-\log 2)^{-\frac{8\pi^2}{g^2_*}}(\frac{b}{2} \log M^2 \overline{\rho^2})^{2N_c} W(\overline{\rho^2})^{\frac{8\pi^2}{g_*^2}},
\end{equation}

\noindent
we can write the 2-loop instanton density as,

\begin{equation}\label{eq:2-loop-instanton-density}
d_{2L}(\rho)=f(\overline{\rho})m^{N_f}\rho^{N_f-5}(\rho \Lambda_c)^{\frac{1}{2}BN_c}.
\end{equation}
From here we find $\overline{\rho^2}$ analogously to the derivation of  (\ref{eq:rho-square-bar-1}). The result can be put in the same form, with

\begin{equation}\label{eq:nu-Banks-Zaks}
\nu=\frac{1}{2}\left(\frac{1}{2}BN_c+N_f-4\right)\ .
\end{equation} 
The minimization of the partition function can now be performed in complete analogy to the derivation of the average instanton number in the pure YM theory and we obtain 

\begin{equation}\label{eq:BZ-num-of-instantons}
\langle N \rangle =V \Lambda_c^4 \left[\Gamma(\nu) (\frac{m}{\Lambda_c})^{N_f} f(\bar{\rho}) (\beta \gamma^2 \nu)^{-\frac{\nu}{2}}\right]^{\frac{2}{2+\nu}} \ .
\end{equation}
\noindent
Comparing to (\ref{eq:avg-instanton-number}), the most notable difference is the appearance of the RG-invariant scale $\Lambda_c$ instead of the IR-divergence scale $\Lambda\simeq\Lambda_{YM}$. Another important thing is that 
$\tilde{\beta}^{2N_c}$ is replaced by $(\frac{m}{\Lambda_c})^{N_f}f(\bar{\rho})$, which renormalizes the 1-loop result (\ref{eq:avg-instanton-number}).
The partition function still has the same form (\ref{eq:RILM-partition-function}) as in the pure YM case, but with new values for $\nu$ and $\langle N \rangle$.

\begin{figure}
\centering
\includegraphics[scale=.7]{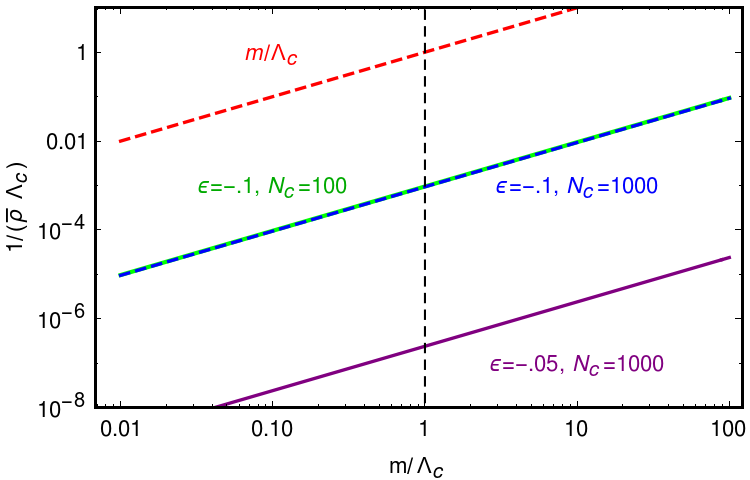}
\caption{\label{fig:BZ-inverse-rho}
The figure shows inverse of $\bar{\rho}$ measured in units of $\Lambda_c$ for various choces of $\epsilon$ and $N_c$ (green, blue and purple). For fixed $\epsilon=-.1$ changing $N_c$ from 100 to 1000 changes $\bar{\rho}$ by less than 2\%. For $\epsilon=-.1$ and $N_c=1000$ changing the parameter a from 1 to .1 has less than .1\% effect. Red dashed line shows fermion mass m in units of $\Lambda_c$.}
\end{figure}

Solving the equations for $\bar{\rho}$ and $N/V$ the way we did in the pure YM case leads us to the results shown in the figure \ref{fig:BZ-inverse-rho}. Crucially, the results are inconsistent with the hypotheses in the sense that $\bar{\rho}^{-1}$ that we find is always smaller than m, i.e. it is more IR than the scale m where we decouple fermions.
This leads us to look for the solution below the energy scale m, where running of the couplings is given by pure YM beta functions \footnote{Equivalently, we may look for solutions using the running coupling defined as a piecewise function, equal to BZ running coupling beyond energy m and equal to matching pure YM running coupling below m.}.
This leads us to consider the equations (\ref{eq:rho-square-bar-1}) and (\ref{eq:avg-instanton-number}) again. 

We know from the subsection \ref{sec:pure-YM} that the solutions for instantons in the pure YM theory are internally consistent, meaning that $\bar{\rho}^{-1}\gg \Lambda_{YM}$. When solving the equations for the BZ theory, since we didn't find any solutions for $\bar{\rho}^{-1} > m$, we additionally have to make sure that the consistency condition $\bar{\rho}^{-1} < m$ is met when using the pure YM running coupling.

\begin{figure}
\centering
\includegraphics[scale=.7]{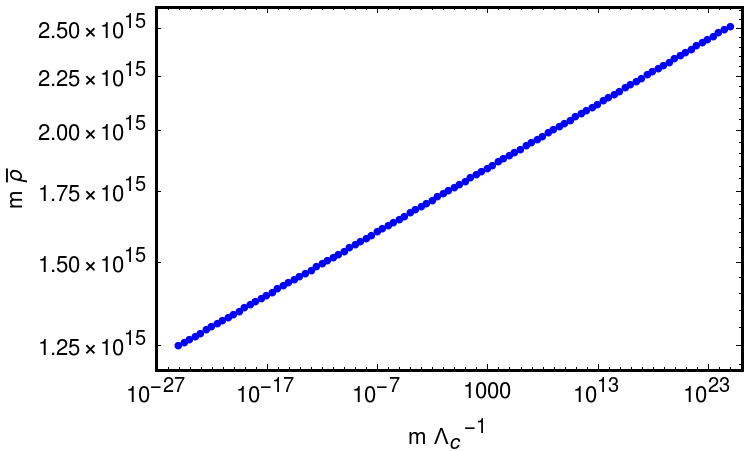}
\includegraphics[scale=.7]{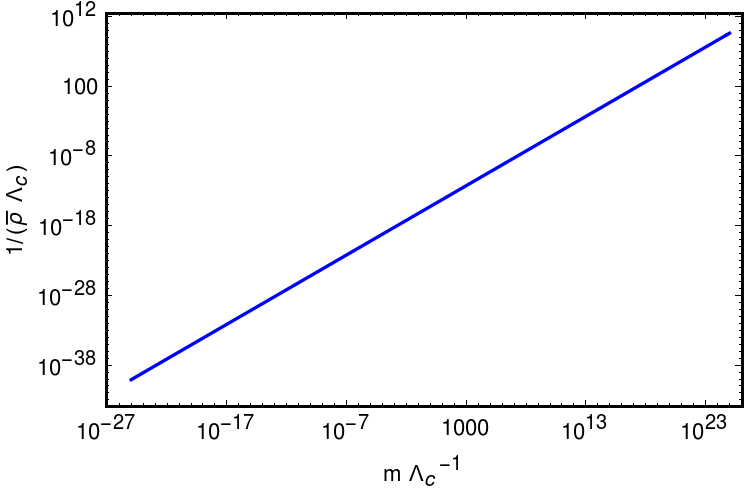}
\caption{\label{fig:consistency}
We take $a=1/10$, $N_c=1000$, and $\epsilon=-1/10$. The top panel shows the instanton scale in the deep IR w.r.t. the fermion mass. The bottom panel shows $\bar{\rho}^{-1}$ as a function of m. Numerical values are predominantly determined by $\epsilon$, with a very mild dependence on $a$ and $N_c$.}
\end{figure}

Our results are shown in the top panel of figure \ref{fig:consistency} which shows the ratio of $m$ to $\bar{\rho}^{-1}$ as a function of $m$ measured in units of $\Lambda_c$. Results for  $\bar{\rho}$ are well within the required consistency range. Bottom panel shows the inverse instanton length as a function of mass, and we can clearly see the power law decrease of the instanton scale as m is taken to zero.

\begin{figure}
\centering
\includegraphics[scale=.7]{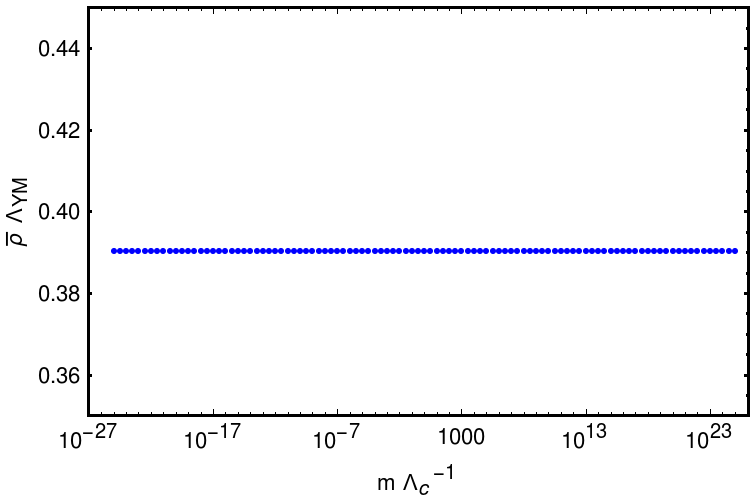}
\caption{\label{fig:BZ-rho-of-MYM}
The figure shows $\bar{\rho}$ in units of $\Lambda_{YM}$ as a function of m. We've fixed $a=1/10,\ N_c=1000$ and $\epsilon=-1/10$.
}
\end{figure}

As an additional consistency check one may study the behavior of $\Lambda_{YM}$. In fact, here it is not an arbitrary number but it is specified by the following one loop  matching conditions 
\begin{equation}
\frac{8\pi^2}{g^2_{YM}(m)}\equiv -\frac{11}{3}N_c\ log( a\frac{\Lambda_{YM}}{m} ) = \frac{8\pi^2}{g^2_{BZ}(m)}=\frac{N_c}{2 \alpha(m)} \ ,
\end{equation}
which  yields, 

\begin{equation}
\Lambda_{YM}=\frac{m}{a} \exp \left( -\frac{3}{22} \frac{1}{\alpha(m)} \right) \ .
\end{equation}
For small enough $\epsilon$ the exponential term is flat as a function of $m$, so the dependence on mass here is essentially linear.  Finally, we can measure $\bar{\rho}$ in units of $\Lambda_{YM}$ and what we find is that it is flat as a function of $m$, taking value $\bar{\rho}=0.390\Lambda_{YM}^{-1}$  {for $a=1/10$, $N_c=1000$ and $\epsilon=-1/10$}. See figure \ref{fig:BZ-rho-of-MYM}.

\vspace{.5cm}

Let us now discuss the instanton energy and the topological susceptibility. Since the couplings are renormalized at the energy scale corresponding to the inverse of the average instanton size, and since the instanton size turns out to be such that they sit well within the pure YM regime, the analysis closely follows the pure YM case. In particular, the partition function again takes the simple form (\ref{eq:RILM-partition-function}) with $\langle N \rangle$ and $\nu$  given by (\ref{eq:avg-instanton-number}) and (\ref{eq:rho-square-bar-1}) respectively. The total energy is given by a sum of the free energy term, $\langle N \rangle$, and the interaction term. The interaction term comes from the derivative of the partition function wrt $\beta=8\pi^2/g^2_{2L}(\bar{\rho})$. This dependency is hidden in $\langle N \rangle$ where it appears in the same form as it did in the pure YM case, which means the interaction energy can again be written as $\langle U_{int} \rangle = \nu \langle N \rangle / (2\beta)$. 
The ratio of the interaction energy to free energy thus  follows the curve shown in figure \ref{fig:YM-energy-ratio}. In fact the shape of that curve changes significantly if 2-loop running is used instead of the 1-loop running and the overall method is more stable when compared to the  QCD case. 
If one fixes $m$, $N_c$ and $a$ (e.g. $m=1/10\ \Lambda_c,\ N_c=1000,\ a=1/10$) one can study $\bar{\rho}$, in units of $\Lambda_{YM}$, as a function of $\epsilon$ and find that it is constant (and in our example) equal to  $\bar{\rho}=0.390\Lambda_{YM}^{-1}$. The reason why  $\bar{\rho}(\epsilon)$ is constant in units of $\Lambda_{YM}$ is related to the fact that $\Lambda_{YM}$ decreases rapidly with decreasing $|\epsilon|$, thus compensating for the rapidly growing $\bar{\rho}$ in units of $\Lambda_c$.

\begin{figure}
\centering
\includegraphics[scale=.7]{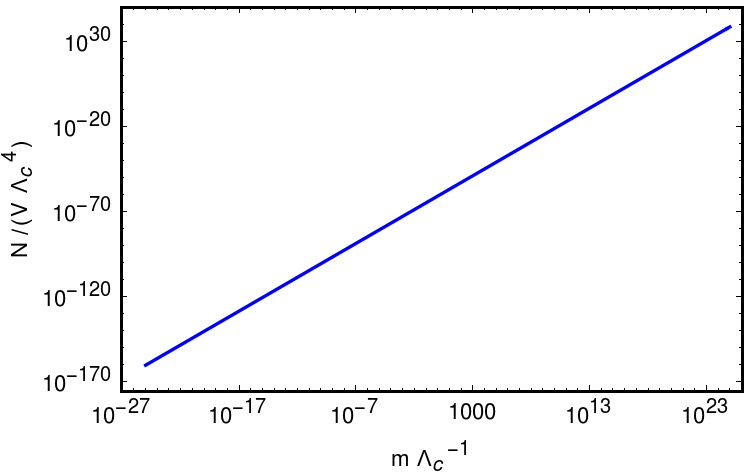}
\includegraphics[scale=.7]{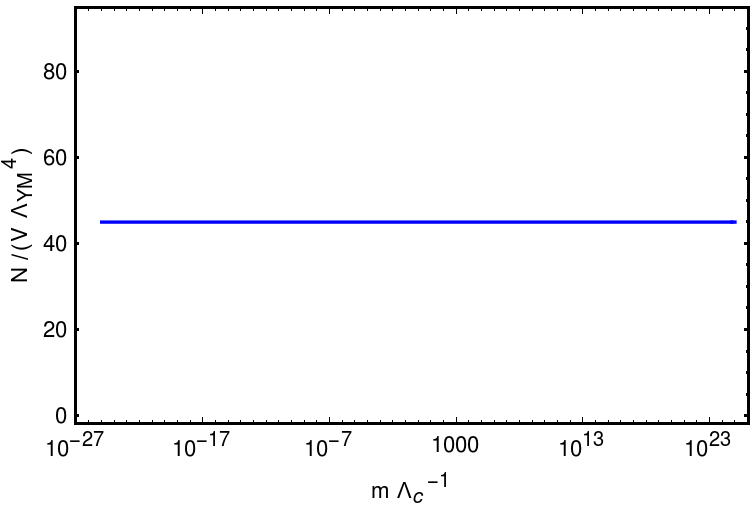}
\caption{\label{fig:BZ-N-over-V}
Top panel shows density of the instantons per unit volume measured in units of $\Lambda_c^{-1}$. Bottom panel shows the same quantity measured in units of $\Lambda_{YM}^{-1}$. The exact value of N/V depends on the fudge factor a. The bottom panel shows $N/(V\Lambda_{YM}^4)=99.268$ obtained for a=1/10. Decreasing a by 1/2 increases N/V by 30\%, and decreasing a by a further factor of 1/2 decreases N/V by additional 23\%.}
\end{figure}

The determination of the topological susceptibility proceeds as described in the previous section, see equation (\ref{eq:RILM-top-suscept-result}). 
As we've discussed above, $\Lambda_{YM} \bar{\rho}$ is essentially $m$- and $\epsilon$-independent (see figure \ref{fig:BZ-rho-of-MYM}). In this sense $N/V$ depends only on the explicit factor $\bar{\rho}^{-4}$. It is then clear that $N/V$ will rapidly decrease with decreasing $m/\Lambda_c$ when measured in units of  $\Lambda_c^{-4}$, but will  be constant if measured in units of $\Lambda_{YM}^{-4}$. This is confirmed in figure \ref{fig:BZ-N-over-V}.

\section{Safe Instantons}
\label{SafeInstantons}

Here we extend the instanton calculus to asymptotically safe quantum field theories starting with the first discovered controllable asymptotically safe four dimensional gauge theory, here dubbed LISA \cite{Litim:2014uca}. 

\subsection{Controllable Instantons in UV-Safe Gauge-Yukawa Theories}

LISA consists of an $SU(N_c)$ gauge field coupled to $N_f$ vector-like fermions and a scalar field. Besides the gauge coupling there are Yukawa couplings and two scalar self-couplings. At 2-loop order the  beta function of the gauge coupling has the LO term exactly as in (\ref{eq:2-loop-betas}), but the cubic term becomes,

\begin{equation}
\left[(25+\frac{26}{3}\epsilon)\alpha-2(\frac{11}{2}+\epsilon)^2\alpha_y\right]\alpha^2,
\end{equation}
where $\alpha_y=y^2 N_c/(4\pi)^2$ and $y$ is the Yukawa coupling. In the Veneziano limit the theory admits a perturbative interacting UV fixed point. At the fixed point value of the gauge coupling, the Yukawa coupling and scalar self-couplings all have values of order $\epsilon$, where $\epsilon$ is again given by (\ref{eq:def-epsilon}) but this time is positive because asymptotic freedom is lost.
To simplify the discussion, herein we neglect the running of the Yukawa coupling. This slightly changes the numerical behavior of the running gauge coupling, but qualitatively the picture of having a running coupling interpolating between a Gaussian FP in the IR and a perturbative, non-Gaussain FP in the UV persists. In particular, substituting the fixed point value of the Yukawa coupling, $\alpha_y^*$, in the above expression for the cubic term leads exactly to the beta function (\ref{eq:2-loop-betas}), with $B=-4\epsilon/3$ and 

\begin{equation}
C=-\frac{2}{3}\frac{57-46\epsilon-8\epsilon^2}{13+\epsilon}\ .
\end{equation}
\noindent
Both $B$ and $C$ being negative, the fixed point appears at the physical value $\alpha_*=B/C>0$. Note that this couldn't have been possible without the inclusion of scalars in the theory, since what was necessary for flipping the sign of the cubic term was the contribution of the Yukawa coupling. From here on it is clear that {running of the gauge coupling in} LISA (in this slightly simplified form) is analogous to {that in} the Banks-Zaks theory, with UV and IR reversed although the dynamics is profoundly different in nature. 

In particular, the 2-loop running gauge coupling is again given by (\ref{eq:2-loop-running}). The difference, of course, is in the fact that the argument of the Lambert function, $z(\rho)$ that is given in (\ref{eq:arg-of-Lambert}), now runs inversely proportional $\rho$   because $B=-4\epsilon/3$ changed sign with respect to the IR interacting fixed point. 

\begin{figure}
\centering
\includegraphics[scale=.7]{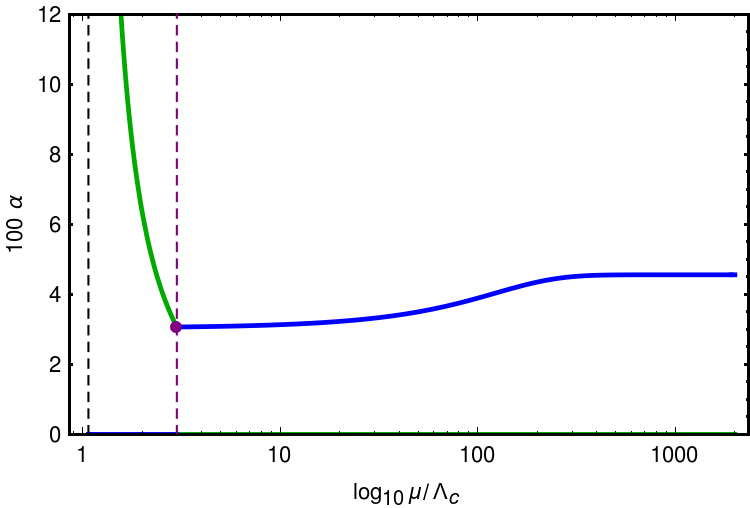} \\
\caption{\label{fig:massive-LISA-running}
The blue line shows the LISA running for $\epsilon=1/10$, and green line corresponds to the pure YM running. Purple dot shows the matching couplings at the fermion mass scale which is given by purple dashed line. Black dashed line is the scale $\Lambda_{YM}$.}
\end{figure}

We include the fermion effects following the previous section. The running coupling behaves according to figure \ref{fig:massive-LISA-running}.
Due to theory being non-interacting in the IR, there are no condensates forming at the fermion mass scale and thus zero modes are taken care of in the same straightforward manner as in the previous section (see \cite{tHooft:1976snw}).

\vspace{.5cm}

Computation of the average instanton size and the density of instantons per unit volume proceeds analogously to the perturbative interacting IR fixed point  theory presented in the previous section. In particular, the 2-loop instanton density is  given by (\ref{eq:2-loop-inst-density}). The main difference is that the argument of the Lambert function, given in (\ref{eq:arg-of-Lambert}), now grows with $\rho$ due to $B=-\nicefrac{4}{3}\ \epsilon<0$. In the remaining terms of (\ref{eq:2-loop-inst-density}) with explicit power-law dependence on $\rho$, $N_f$ dominates over $\nicefrac{1}{2}B N_c$, so the fact that B changes sign here is irrelevant.

From this 2-loop density one can obtain the effective 2-loop density $\mu(\rho)$  in a similar way as done in the IR interacting case. It will therefore again lead to the expression (\ref{eq:naive-optimal-mu}). Note that $\beta$ is just shorthand for $8\pi^2/g^2$ so it doesn't change sign wrt the IR interacting theory, in fact, we still have a Gaussian suppression of the IR instantons. 
The expectation value of $\rho$ that we get is   (\ref{eq:rho-square-bar-1}) with $\nu$ given by (\ref{eq:nu-Banks-Zaks}). Here $\beta$ is still positive, so $\nu$ has to be positive too if $\bar{\rho}$ is to be positive. In fact, $\nicefrac{1}{2}BN_c=-\nicefrac{4}{6}\epsilon N_c$, whereas $N_f=(\nicefrac{11}{2}+\epsilon)N_c\simeq \nicefrac{11}{2} N_c$, and thus $\nu$ is clearly positive both in the  BZ and LISA theories.

\begin{figure}
\centering
\includegraphics[scale=.7]{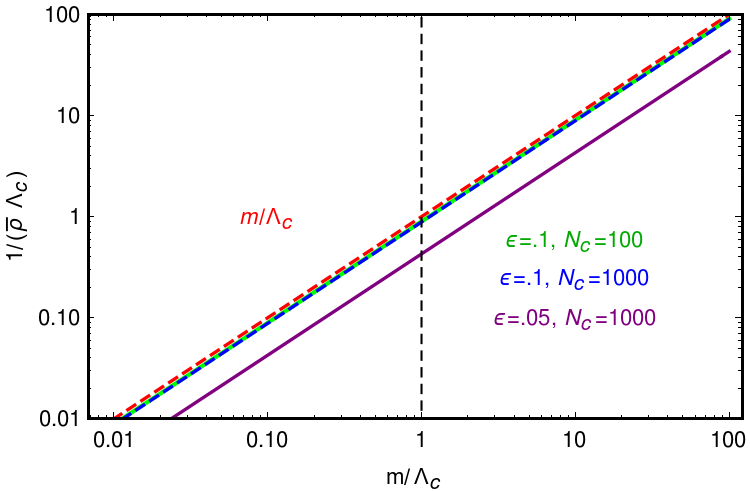}
\caption{\label{fig:LISA-inconsistent-solns}
The figure shows solutions for $\bar{\rho}$, obtained using LISA beta functions, for various choices of $\epsilon$ and $N_c$ (green, blue and purple). Red dashed line shows the fermion mass. }
\end{figure}

\begin{figure}
\centering
\includegraphics[scale=.7]{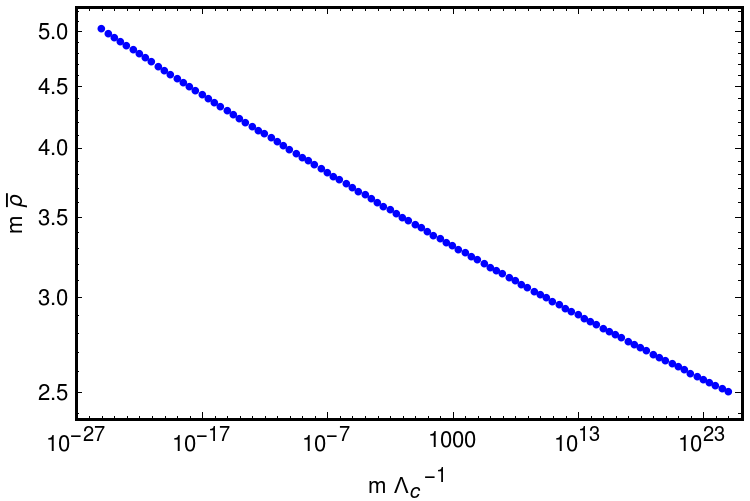}
\includegraphics[scale=.7]{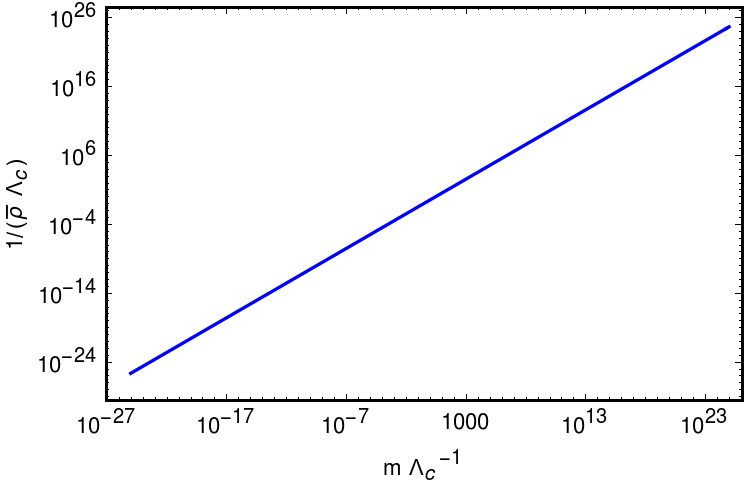}
\caption{\label{fig:LISA-instanton-scale}
Top panel shows ratio $m/\bar{\rho}^{-1}$ and bottom panel shows $\bar{\rho}^{-1}$ as functions of m. In both figures a=1/10, $N_c=1000$ and $\epsilon=1/10$.}
\end{figure}

Finally, $\langle N \rangle$ takes the same form (\ref{eq:BZ-num-of-instantons}) as in the  BZ case with $\nu$, $\beta$ and $f(\rho)$ appropriately modified.
Solving the equations for $\langle N \rangle$ and $\bar{\rho}$ numerically, we find results similar to the  BZ instantons.
In particular, using the LISA beta functions we do find solutions for $\bar{\rho}$, with the instanton scale still smaller than the  fermion mass, which means that the results are not consistent (see figure \ref{fig:LISA-inconsistent-solns}). We then solve the equations using the pure YM beta functions and find solutions with the instanton scale in the window between the IR $\Lambda_{YM}$ scale (which is expected from \ref{sec:pure-YM}), and the fermion decoupling scale (which is a nontrivial consistency check). For the results in LISA see figure \ref{fig:LISA-instanton-scale}.

It is of some interest to compare figures \ref{fig:LISA-instanton-scale} and \ref{fig:consistency}. One interesting feature is that, for the chosen parameters, the instanton scale is about one order of magnitude smaller than the fermion mass in the case of LISA, but it is more than 16 orders of magnitude smaller than the fermion mass in the BZ theory.  Here the difference arises because in the infrared LISA is free rather than interacting.  Further, in both cases the instanton scale $\bar{\rho}^{-1}$ always has to lie below the mass scale $m$, which explains the fact that the lines in the bottom panels have the same tendency to grow with $m$. Finally, from the top panels we see that the ratio $m/ \bar{\rho}^{-1}$ grows with $m$ in BZ theory but decreases in LISA. In fact, the BZ theory is non-interacting in the UV and the higher the energies at which we decouple fermions the higher the IR instanton scale seems to be. This pattern is found in the LISA case as well. This behavior is related to how  $\beta=8\pi^2/g^2$ enters the equations (\ref{eq:avg-instanton-number}) and (\ref{eq:rho-square-bar-1}).

\begin{figure}
\centering
\includegraphics[scale=.7]{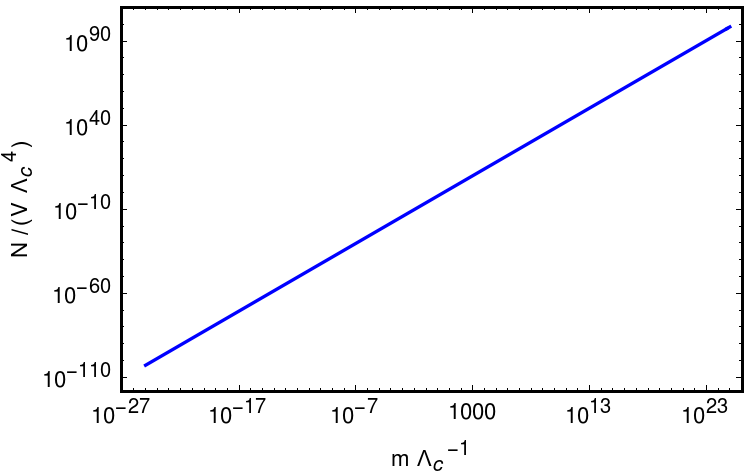}
\caption{\label{fig:LISA-N-over-V}
The figure shows density of the instantons per unit volume measured in units of $\Lambda_c^{-4}$. }
\end{figure}

We can study $\bar{\rho}$ in units of $\Lambda_{YM}$ as a function of $m$, but it is clear that the results are described by figure \ref{fig:BZ-rho-of-MYM}, i.e. in units of $\Lambda_{YM}$ the solution reproduces the BZ result. The same holds for $\langle N \rangle /V$ in units of $\Lambda_{YM}^{-4}$ which is shown in bottom panel of figure \ref{fig:BZ-N-over-V}. For the instanton density in units of $\Lambda_c^{-4}$ see figure \ref{fig:LISA-N-over-V}.

 Because of the perturbative nature of the UV interacting fixed point we have been able to extend the instanton calculus to controllable asymptotically safe quantum field theories. 
  
\subsection{Safe QCD Instantons, the large $N_f$ story}

 In the LISA theory \cite{Litim:2014uca} elementary scalars and their induced Yukawa interactions crucially help taming the ultraviolet behaviour of the overall gauge-Yukawa theory.  Scalars, however, are not needed at finite number of colours and very large number of flavours for non-abelian gauge-fermion theories as reviewed and further analysed in \cite{Antipin:2017ebo}. 
 
Consider an $SU(N_c)$ gauge theory with $N_f$ fermions transforming according to a given representation of the gauge group. We consider the theory for a number of flavours above which asymptotic freedom is lost, i.e. $N_f^{AF} > 11C_G/(4T_R)$, where the first coefficient of the beta function changes sign. Although we don't not need to specify the fermion representation we will consider here  the fundamental representation for which the relevant group theory coefficients  are  $C_G=N_c$, $C_R=(N_c^2-1)/2N_c$ and $T_R=1/2$. At one loop order the theory is simultaneously  free in the infrared  (non-abelian QED) and trivial, meaning that  the only  way to take the continuum limit (i.e. sending the Landau pole induced cutoff to infinity) is for the theory to become non-interacting. At two-loops Caswell \cite{Caswell:1974gg} demonstrated that   an UV interacting fixed point  (asymptotic safety) cannot arise near the loss of asymptotic freedom implying that  safety can only occur above a certain critical number of flavours.  This possibility has been (re)investigated in  \cite{Antipin:2017ebo} at large  $N_f$  and fixed number of colours for which the  beta function is given by \cite{PalanquesMestre:1983zy,Gracey:1996he,Holdom:2010qs,Pica:2010xq}.  
 
\begin{equation}
\beta(A)=\frac{2A}{3}\left( 1+\sum_{i=1}^{\infty} \frac{H_i(A)}{N_f^i} \right)\ ,
\end{equation}
where we defined the following large $N_f$ normalized coupling

\begin{equation}
A=\frac{N_f}{8\pi^2}g^2 \ .
\end{equation}
The functions $H_i(A)$ come about by resumming an infinite set of Feynman diagrams at fixed order $i$ in the large-$N_f$ expansion \cite{PalanquesMestre:1983zy,Gracey:1996he}. 
Most importantly, already to first order in large-$N_f$ there is a fixed point,

\begin{equation}
A^* = 3 - e^{\left(-8 \frac{N_f}{N_c} + 18.49 - 5.26 \frac{N_c^2 - 1}{2 N_c^2}\right)} \ .
\end{equation}

We now attempt to approximate the overall behaviour of the beta function in order to  estimate the instanton properties for this theory. Let us therefore write the 1-loop  running as,

\begin{equation}
\beta_{1L}\equiv\frac{8\pi^2}{g^2}= -b \log(a_{LP} \Lambda_{LP} \rho)\ ,
\end{equation}
and the 2-loop  running as,
\begin{equation}
\beta_{2L}= \beta_{1L} + \frac{b'}{b}\log\beta_{1L} \ ,
\end{equation}

where the 1-loop and 2-loop coefficients are given in (\ref{eq:1Lbeta}) and (\ref{eq:naive-2-loop-running}). We fix the fudge factor $a_{LP}$ by requiring that the 2-loop running $g^2$ matches the UV FP value of $g^2$ at the 1-loop divergent scale $\Lambda_{LP}$. 

\begin{figure}
\includegraphics[scale=.7]{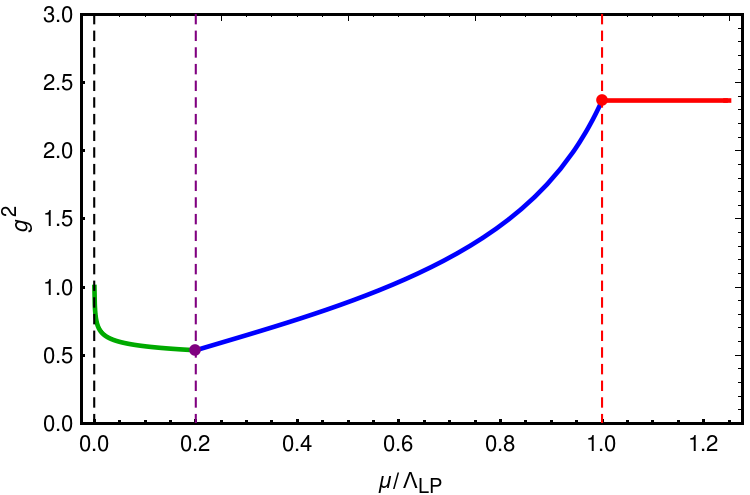}
\caption{\label{fig:large-nf-couplings}
Flow of the coupling in the mass-deformed large-$N_f$ UV conformal window. Pure Yang-Mills 1-loop running is shown in green. Matching to 2-loop QCD running, shown in blue, is at the fermion mass scale given by the purple dashed line. We match 2-loop QCD running to the UV FP value, shown in red, at the 1-loop UV divergence scale $\Lambda_{LP}$ shown in red dashed line.
}
\end{figure}

\begin{figure}
\includegraphics[scale=.7]{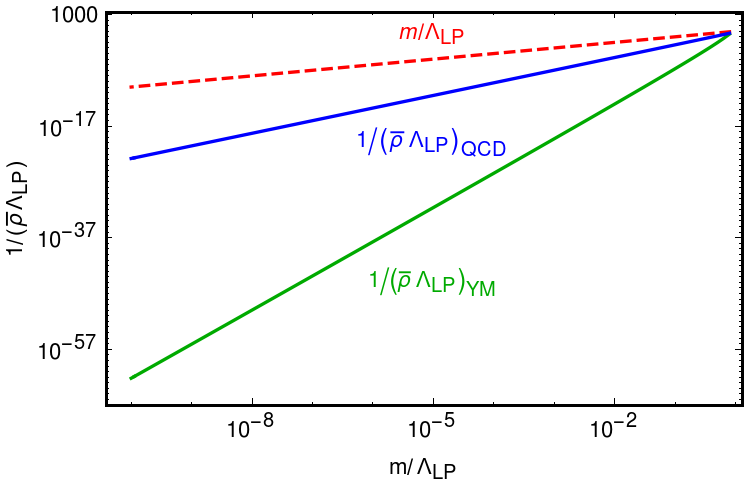}
\caption{\label{fig:large-nf-consistency}
Red dashed line shows the fermion mass scale. Solving for the $1/\bar{\rho}$ using 2-loop QCD beta functions we find the solutions shown in blue. This puts instantons below the fermion decoupling scale which makes the solutions inconsistent. Next we solve for the $1/\bar{\rho}$ using 2-loop pure YM beta functions and we find the solutions shown in green. We see that the instanton scale is still below the fermion mass scale and is thus consistent.
}
\end{figure}

Let us consider the  specific examples $N_f=100$, $N_c=3$ for which $a_{LP}=1.189$. The running couplings are given in  figure \ref{fig:large-nf-couplings}. There are three energy windows, the lowest one is below the fermion mass scale, the intermediate one runs up to the 1-loop divergent scale $\Lambda_{LP}$ (Landau pole), and the highest one above the scale $\Lambda_{LP}$. In the highest energy window we don't consider the RG running, but instead keep the coupling constant since it reaches the fixed point value at $\Lambda_{LP}$.

We only consider fermion masses $m<\Lambda_{LP}$ and in figure \ref{fig:large-nf-consistency} we plot the results for the inverse instanton size (in units of $\Lambda_{LP}$) in the lower two energy windows. We see that using the naive running couplings the instanton scale turns out to be very small, similarly to what we found in BZ and LISA. This is perhaps not surprising given that  we adopted a naive setup that makes use of a rough 2-loop approximation. 

\section{Conclusions}
\label{conclusions}

We investigated the instanton dynamics for fundamental field theories featuring either an asymptotically safe or free dynamics. In order to make the work self-contained we provided a brief review of the role  of instantons for YM and QCD dynamics including the limitations of the instanton calculus. 
Within the asymptotically free scenario we  ventured in the perturbative regime of the QCD conformal window. Here we determined, by extending the calculus to two-loops, the number of instantons per unit volume as function of a common fermion mass.

We then extended the instanton calculus to the case of  controllable asymptotically safe  theories. Here the non-trivial UV dynamics demands  the immediate use of higher order results. As for the conformal window case we determined the fermion mass dependence of the instanton density. We further discussed the finite number of colours and large number of flavours limit.

In the future one will extend the instanton calculus in order to cover a wider range of number of flavors within the calculable regime of UV and IR conformal windows.  The ambitious goal is to determine to which extent the instanton dynamics is responsible for the loss of conformality once the number of flavors drops below a certain critical value for which either UV or IR conformality is lost in the respective safe or free conformal windows. 
  
\acknowledgments 
The work of F.S. is partially supported by the Danish National Research Foundation under grant DNRF:90.

\bibliographystyle{ieeetr} 
\bibliography{safe_instantons_biblio}

\begin{thebibliography}{10}

\bibitem{Wilson:1971bg}
K.~G. Wilson, ``{Renormalization group and critical phenomena. 1.
  Renormalization group and the Kadanoff scaling picture},'' {\em Phys. Rev.},
  vol.~B4, pp.~3174--3183, 1971.

\bibitem{Wilson:1971dh}
K.~G. Wilson, ``{Renormalization group and critical phenomena. 2. Phase space
  cell analysis of critical behavior},'' {\em Phys. Rev.}, vol.~B4,
  pp.~3184--3205, 1971.

\bibitem{Gross:1973ju}
D.~J. Gross and F.~Wilczek, ``{Asymptotically Free Gauge Theories. 1},'' {\em
  Phys. Rev.}, vol.~D8, pp.~3633--3652, 1973.

\bibitem{Politzer:1973fx}
H.~D. Politzer, ``{Reliable Perturbative Results for Strong Interactions?},''
  {\em Phys. Rev. Lett.}, vol.~30, pp.~1346--1349, 1973.

\bibitem{Litim:2014uca}
D.~F. Litim and F.~Sannino, ``{Asymptotic safety guaranteed},'' {\em JHEP},
  vol.~12, p.~178, 2014.

\bibitem{Litim:2015iea}
D.~F. Litim, M.~Mojaza, and F.~Sannino, ``{Vacuum stability of asymptotically
  safe gauge-Yukawa theories},'' {\em JHEP}, vol.~01, p.~081, 2016.

\bibitem{Sannino:2014lxa}
F.~Sannino and I.~M. Shoemaker, ``{Asymptotically Safe Dark Matter},'' {\em
  Phys. Rev.}, vol.~D92, no.~4, p.~043518, 2015.

\bibitem{Abel:2017ujy}
S.~Abel and F.~Sannino, ``{Radiative symmetry breaking from interacting UV
  fixed points},'' {\em Phys. Rev.}, vol.~D96, no.~5, p.~056028, 2017.

\bibitem{Abel:2017rwl}
S.~Abel and F.~Sannino, ``{Framework for an asymptotically safe Standard Model
  via dynamical breaking},'' {\em Phys. Rev.}, vol.~D96, no.~5, p.~055021,
  2017.

\bibitem{Pelaggi:2017wzr}
G.~M. Pelaggi, F.~Sannino, A.~Strumia, and E.~Vigiani, ``{Naturalness of
  asymptotically safe Higgs},'' {\em Front.in Phys.}, vol.~5, p.~49, 2017.

\bibitem{Mann:2017wzh}
R.~Mann, J.~Meffe, F.~Sannino, T.~Steele, Z.-W. Wang, and C.~Zhang,
  ``{Asymptotically Safe Standard Model via Vectorlike Fermions},'' {\em Phys.
  Rev. Lett.}, vol.~119, no.~26, p.~261802, 2017.

\bibitem{Pelaggi:2017abg}
G.~M. Pelaggi, A.~D. Plascencia, A.~Salvio, F.~Sannino, J.~Smirnov, and
  A.~Strumia, ``{Asymptotically Safe Standard Model Extensions},'' 2017.

\bibitem{Bond:2017wut}
A.~D. Bond, G.~Hiller, K.~Kowalska, and D.~F. Litim, ``{Directions for model
  building from asymptotic safety},'' {\em JHEP}, vol.~08, p.~004, 2017.

\bibitem{Banks:1981nn}
T.~Banks and A.~Zaks, ``{On the Phase Structure of Vector-Like Gauge Theories
  with Massless Fermions},'' {\em Nucl. Phys.}, vol.~B196, pp.~189--204, 1982.

\bibitem{Sannino:2009za}
F.~Sannino, ``{Conformal Dynamics for TeV Physics and Cosmology},'' {\em Acta
  Phys. Polon.}, vol.~B40, pp.~3533--3743, 2009.

\bibitem{Pica:2017gcb}
C.~Pica, ``{Beyond the Standard Model: Charting Fundamental Interactions via
  Lattice Simulations},'' {\em PoS}, vol.~LATTICE2016, p.~015, 2016.

\bibitem{PalanquesMestre:1983zy}
A.~Palanques-Mestre and P.~Pascual, ``{The 1/$N^-$f Expansion of the $\gamma$
  and Beta Functions in {QED}},'' {\em Commun. Math. Phys.}, vol.~95, p.~277,
  1984.

\bibitem{Gracey:1996he}
J.~A. Gracey, ``{The QCD Beta function at O(1/N(f))},'' {\em Phys. Lett.},
  vol.~B373, pp.~178--184, 1996.

\bibitem{Holdom:2010qs}
B.~Holdom, ``{Large N flavor beta-functions: a recap},'' {\em Phys. Lett.},
  vol.~B694, pp.~74--79, 2011.

\bibitem{Pica:2010xq}
C.~Pica and F.~Sannino, ``{UV and IR Zeros of Gauge Theories at The Four Loop
  Order and Beyond},'' {\em Phys. Rev.}, vol.~D83, p.~035013, 2011.

\bibitem{Shrock:2013cca}
R.~Shrock, ``{Study of Possible Ultraviolet Zero of the Beta Function in Gauge
  Theories with Many Fermions},'' {\em Phys. Rev.}, vol.~D89, no.~4, p.~045019,
  2014.

\bibitem{Antipin:2017ebo}
O.~Antipin and F.~Sannino, ``{Conformal Window 2.0: The Large $N_f$ Safe
  Story},'' 2017.

\bibitem{Intriligator:2003jj}
K.~A. Intriligator and B.~Wecht, ``{The Exact superconformal R symmetry
  maximizes a},'' {\em Nucl. Phys.}, vol.~B667, pp.~183--200, 2003.

\bibitem{Hofman:2008ar}
D.~M. Hofman and J.~Maldacena, ``{Conformal collider physics: Energy and charge
  correlations},'' {\em JHEP}, vol.~05, p.~012, 2008.

\bibitem{Intriligator:2015xxa}
K.~Intriligator and F.~Sannino, ``{Supersymmetric asymptotic safety is not
  guaranteed},'' {\em JHEP}, vol.~11, p.~023, 2015.

\bibitem{Martin:2000cr}
S.~P. Martin and J.~D. Wells, ``{Constraints on ultraviolet stable fixed points
  in supersymmetric gauge theories},'' {\em Phys. Rev.}, vol.~D64, p.~036010,
  2001.

\bibitem{Seiberg:1994pq}
N.~Seiberg, ``{Electric - magnetic duality in supersymmetric nonAbelian gauge
  theories},'' {\em Nucl. Phys.}, vol.~B435, pp.~129--146, 1995.

\bibitem{Bajc:2016efj}
B.~Bajc and F.~Sannino, ``{Asymptotically Safe Grand Unification},'' {\em
  JHEP}, vol.~12, p.~141, 2016.

\bibitem{Bajc:2017xwx}
B.~Bajc, N.~A. Dondi, and F.~Sannino, ``{Safe SUSY},'' 2017.

\bibitem{Shifman:2012zz}
M.~Shifman, {\em {Advanced topics in quantum field theory.}}
\newblock Cambridge, UK: Cambridge Univ. Press, 2012.

\bibitem{Shuryak:1988ck}
E.~V. Shuryak, ``{The QCD vacuum, hadrons and the superdense matter},'' {\em
  World Sci. Lect. Notes Phys.}, vol.~71, pp.~1--618, 2004.
\newblock [World Sci. Lect. Notes Phys.8,1(1988)].

\bibitem{Coleman1988}
S.~Coleman, {\em Aspects of symmetry: selected Erice lectures}.
\newblock Cambridge University Press, 1988.

\bibitem{Schafer:1996wv}
T.~Schäfer and E.~V. Shuryak, ``{Instantons in QCD},'' {\em Rev. Mod. Phys.},
  vol.~70, pp.~323--426, 1998.

\bibitem{Vainshtein:1981wh}
A.~I. Vainshtein, V.~I. Zakharov, V.~A. Novikov, and M.~A. Shifman, ``{ABC's of
  Instantons},'' {\em Sov. Phys. Usp.}, vol.~25, p.~195, 1982.
\newblock [Usp. Fiz. Nauk136,553(1982)].

\bibitem{Belavin:1975fg}
A.~A. Belavin, A.~M. Polyakov, A.~S. Schwartz, and {\relax Yu}.~S. Tyupkin,
  ``{Pseudoparticle Solutions of the Yang-Mills Equations},'' {\em Phys.
  Lett.}, vol.~59B, pp.~85--87, 1975.

\bibitem{tHooft:1976snw}
G.~'t~Hooft, ``{Computation of the Quantum Effects Due to a Four-Dimensional
  Pseudoparticle},'' {\em Phys. Rev.}, vol.~D14, pp.~3432--3450, 1976.
\newblock [Erratum: Phys. Rev.D18,2199(1978)].

\bibitem{Bernard:1977nr}
C.~W. Bernard, N.~H. Christ, A.~H. Guth, and E.~J. Weinberg, ``{Instanton
  Parameters for Arbitrary Gauge Groups},'' {\em Phys. Rev.}, vol.~D16,
  p.~2967, 1977.

\bibitem{Bernard:1979qt}
C.~W. Bernard, ``{Gauge Zero Modes, Instanton Determinants, and QCD
  Calculations},'' {\em Phys. Rev.}, vol.~D19, p.~3013, 1979.

\bibitem{Shifman:1979uw}
M.~A. Shifman, A.~I. Vainshtein, and V.~I. Zakharov, ``{Instanton Density in a
  Theory with Massless Quarks},'' {\em Nucl. Phys.}, vol.~B163, pp.~46--56,
  1980.

\bibitem{Diakonov:1985eg}
D.~Diakonov and V.~{\relax Yu}. Petrov, ``{A Theory of Light Quarks in the
  Instanton Vacuum},'' {\em Nucl. Phys.}, vol.~B272, pp.~457--489, 1986.

\bibitem{Callan:1977gz}
C.~G. Callan, Jr., R.~F. Dashen, and D.~J. Gross, ``{Toward a Theory of the
  Strong Interactions},'' {\em Phys. Rev.}, vol.~D17, p.~2717, 1978.

\bibitem{Shuryak:1982dp}
E.~V. Shuryak, ``{The Role of Instantons in Quantum Chromodynamics. 2. Hadronic
  Structure},'' {\em Nucl. Phys.}, vol.~B203, pp.~116--139, 1982.

\bibitem{Diakonov:1983hh}
D.~Diakonov and V.~{\relax Yu}. Petrov, ``{Instanton Based Vacuum from Feynman
  Variational Principle},'' {\em Nucl. Phys.}, vol.~B245, pp.~259--292, 1984.

\bibitem{Caswell:1974cj}
W.~E. Caswell and F.~Wilczek, ``{On the Gauge Dependence of Renormalization
  Group Parameters},'' {\em Phys. Lett.}, vol.~49B, pp.~291--292, 1974.

\bibitem{Schafer:2002af}
T.~Schäfer, ``{Instantons in QCD with many colors},'' {\em Phys. Rev.},
  vol.~D66, p.~076009, 2002.

\bibitem{Ryttov:2010iz}
T.~A. Ryttov and R.~Shrock, ``{Higher-Loop Corrections to the Infrared
  Evolution of a Gauge Theory with Fermions},'' {\em Phys. Rev.}, vol.~D83,
  p.~056011, 2011.

\bibitem{Ryttov:2016ner}
T.~A. Ryttov and R.~Shrock, ``{Infrared Zero of $\beta$ and Value of $\gamma_m$
  for an SU(3) Gauge Theory at the Five-Loop Level},'' {\em Phys. Rev.},
  vol.~D94, no.~10, p.~105015, 2016.

\bibitem{Sannino:2010ca}
F.~Sannino, ``{Mass Deformed Exact S-parameter in Conformal Theories},'' {\em
  Phys. Rev.}, vol.~D82, p.~081701, 2010.

\bibitem{Caswell:1974gg}
W.~E. Caswell, ``{Asymptotic Behavior of Nonabelian Gauge Theories to Two Loop
  Order},'' {\em Phys. Rev. Lett.}, vol.~33, p.~244, 1974.

\end{thebibliography}

\end{document}